\documentclass[10pt]{article}
\usepackage[titletoc,title,toc]{appendix}
\usepackage[margin=1in]{geometry}
\usepackage{graphicx}
\usepackage{amssymb}
\usepackage{epstopdf}
\usepackage{amsmath}
\usepackage{color}
\usepackage{hyperref}
\usepackage{mathrsfs}
\usepackage{varioref}
\usepackage[normalem]{ulem}
\usepackage[active]{srcltx}
\expandafter\ifx\csname package@font\endcsname\relax\else
\expandafter\expandafter
\expandafter\usepackage
\expandafter\expandafter
\expandafter{\csname package@font\endcsname}%
\fi
\hypersetup{
    bookmarks=false,         
    pdfstartview={FitH},    
    colorlinks=true,       
    linkcolor=red,          
    citecolor=blue,        
    filecolor=blue,      
    urlcolor=blue           
}
\DeclareFontFamily{OT1}{pzc}{}
\DeclareFontShape{OT1}{pzc}{m}{it}%
             {<-> s * [0.900] pzcmi7t}{}
\DeclareMathAlphabet{\mathscr}{OT1}{pzc}%
                                 {m}{it}
\labelformat{section}{Section #1} 
\labelformat{subsection}{Section #1} 
\labelformat{equation}{Eq.(#1)} 
\labelformat{figure}{Fig.~#1} 
\labelformat{subfigure}{Fig.~\thefigure#1} 
\labelformat{table}{Tab.~#1} 
\newcommand{\be}{\begin{equation}}
\newcommand{\ee}{\end{equation}}
\newcommand{\bea}{\begin{eqnarray}}
\newcommand{\eea}{\end{eqnarray}}

\newcommand{\cosec}{\operatorname{cosec}}
\newcommand{\cosech}{\operatorname{cosech}}

\begin{document}

\title{{Quantum Evolution Leading to Classicality:\newline A Concrete Example}}
\author{ 
	{\bf {\normalsize Kinjalk Lochan$^a$,}$
	$\thanks{E-mail: kinjalk@iucaa.ernet.in}} \, 
	{\bf {\normalsize Krishnamohan Parattu$^a$,}$
	$\thanks{E-mail: krishna@iucaa.ernet.in}} \, and
	{\bf {\normalsize T. Padmanabhan$^a$}$
	$\thanks{E-mail: paddy@iucaa.ernet.in}}\\
	{\normalsize $^a$IUCAA, Post Bag 4, Ganeshkhind,}
	\\{\normalsize Pune University Campus, Pune 411 007, India}
	\\[0.3cm]
}

\maketitle

\begin{abstract}
Can certain degrees of freedom of a closed physical system, described by a time-independent Hamiltonian, become more and more classical as 
they evolve from some state? This question is important because our universe seems to have done just that! We construct an explicit, simple,
example of such a system with just two degrees of freedom, of which one achieves `spontaneous classicalization'. It starts from a quantum state
and under the usual Hamiltonian evolution, becomes more and more classical (in a well-defined manner in terms of the Wigner function) as time 
progresses. This is achieved without the usual procedures of integrating out a large number of environmental degrees of freedom or conventional
decoherence. We consider different ranges of parameter space and identify the conditions under which spontaneous classicalization occurs in our
model. The mutual interaction between the sub-systems of a larger system can indeed drive some of the subsystems to a classical configuration,
with a phase space trajectory of evolution. We also argue that the results of our toy model may well be general characteristics of certain class
of interacting systems. Several implications are discussed.
\end{abstract}

\section{Introduction}

The issue of quantum to classical transition of physical systems has attracted great amount of interest since the inception of quantum theory. With the triumph of the quantum theory over the years, we have come to accept the idea that any system --- at a fundamental level --- will be described by the principles of  quantum theory. The idea of deterministic evolution in classical physics has to give way to the probabilistic interpretation of the quantum theory in which systems actually evolve according to probabilistic quantum laws.
On the other hand, we also know for a fact that the classical equations of motion (without any probabilistic interpretation) also seem to work, at least for macroscopic objects in nature.  This contrast, in one form or another, has intrigued the physics community ever since the successful arrival of the quantum theory on the scene.
Standard quantum theory suggests that particles like electrons, atoms, molecules etc --- which are the building blocks of larger chunks of matter --- should be subject to the quantum laws. We can indeed  observe and verify the quantum predictions at the corresponding microscopic scales. But the macroscopic objects, built from the very same entities which obey quantum laws  appear to evolve in a deterministic manner.
Why is that even though the nature is fundamentally probabilistic, the macroscopic objects  around us seem to be behave in a deterministic manner for all practical purposes? How can these two --- seemingly  orthogonal notions --- be reconciled? 

This discomfort has  led to intense debates triggered by different perspectives as to which notions one should (or should not) adhere to and even attempts to modify the quantum theory.  There are proposals in quantum theory, \cite{Zurek:1981xq, Zurek:1982ii, Joos:1984uk}, which try to explain why macroscopic objects tend to satisfy the classical notion of determinism as an emergent and effective phenomena. But there are also counter-arguments \cite{Adler:2001us,  Schlosshauer:2003zy} exploring the limitations of such attempts.

One popular approach for explaining the classical behavior of macroscopic objects is that of {\it decoherence} \cite{Zurek:1981xq, Zurek:1982ii, Joos:1984uk}, which suggests that what we perceive as classicality is in fact an effective
attribute of a subsystem which we are temporarily concerned about. The subsystem, in reality, is in interaction with a large number of other quantum degrees of freedom which we chose to ignore. This choice to ignore all other degrees of freedom (except those of our subsystem), is the root cause of an effective evolution for the subsystem which is indistinguishable, for all practical purposes, from a deterministic evolution. In this approach, the apparent classicality of the macroscopic objects (or microscopic objects interacting with an apparatus) is related to the instability of majority of states in the Hilbert space, caused by the interaction of the subsystem with the environment (including the apparatus). Only very few of the states remain stable in the course of interaction, which serve as the {\it pointer states} of the composite system. The system can remain in the mixture of the pointer states, but the quantum superposition of pointer states leads to unstable states \cite{Zurek:1981xq}. Thus, the environment induces the  selection of certain privileged states, depending upon the details of the environment and of its interaction with the system, which the system can ultimately end up in. All this happens within a {\it decoherence time} which is much smaller than any other dynamical time scale in the system.

This approach, as a possible solution to the problem of classicality observed in macroscopic systems, seems to be prima facie satisfactory. However, there are  debates regarding whether it is indeed the complete answer \cite{Adler:2001us,  Schlosshauer:2003zy}. Is it sufficient to end up with classical expectation values being dominant in order to term the system as classical? Moreover, this approach depends heavily upon the contribution of the environment  we chose to ignore. This makes it inapplicable for the cases in which the classical behavior is sought for closed systems without any environment. That is, this scheme is not applicable for a closed system which  starts in a quantum state and gradually evolves into one with a classical behavior. We know at least one such system, namely our universe, which, to the best of our understanding, has successfully done so. Although there are attempts to apply decoherence mechanism to the early universe, with a selection of a part of the universe as the system of interest and everything else as environment \cite{Padmanabhan:1989rm,Halliwell:1989vw,Kiefer:2008ku, Kiefer:2006je, Kiefer:1998qe, Polarski:1995jg}, it is still not fully clear if this selection of system and environment is very natural or will necessarily end up in yielding classicality \cite{Martin:2012pe, Das:2013qwa}. 

Other competing suggestions \cite{Bassi:2003gd, Bassi:2012bg} make use of interaction of the system with a plausible stochastic field to induce the emergence of classicality.  However, the origin of the stochastic field or its contribution to the evolution of universe are quite ambiguous. It is also not clear whether such a scheme is applicable for a closed system. A common entity in both the approaches seems to be the interaction of the system with some ``other degrees of freedom", which in one case is the large number of environmental degrees of freedom and in the other is an unknown stochastic field. In a sense, both the approaches rely on the evolution being affected by unknown external degrees of freedom which are larger in number than those in the system we are focusing on. 
In the context of quantum cosmology, the role of decoherence has traditionally been explored in great detail \cite{Kiefer:2008ku, Kiefer:2006je, Kiefer:1998qe, Polarski:1995jg,Halliwell:1989vw, Kiefer:1987ft, Padmanabhan:1989rm} with only some recent attempts invoking other approaches for the issue \cite{Martin:2012pe, Das:2013qwa}.

To clarify the distinction between the quantum behavior and the expected classical behavior of a system, it is more convenient to consider the behavior in phase space (i.e, $x-p$ space) than in the configuration space. One way to see this is to note that the uncertainty principle is a relation between conjugate variables and hence is more easily dealt with in phase space than in the configuration space. Classically, we can set the initial condition for a system to be some point in phase space, i.e, some particular pair of values of $(x,p)$. Then, under the classical equations of motion, this point moves in the phase space on a trajectory. Given an initial condition, the system will be found on a point on the trajectory at any later time. The situation is very different in the quantum domain. Unlike in classical case, we cannot set the initial condition as a point in phase space as it is forbidden by the uncertainty principle. The quantum description is formulated in terms of a distribution, given by the wavefunction, which in the phase space, is manifested as Wigner quasi-probability distribution (quasi since it is not necessarily always positive definite) or the Wigner function \cite{Wigner:1932}. The Wigner function provides an equivalent formulation of quantum theory, in the phase space. But since, in general, the Wigner function cannot always be seen as a classical distribution in phase space owing to possible regions of its negative values, this presents another possible point of departure from classical physics.  

Once we have defined the initial state as such a distribution in phase space, the evolution equation derived from the Schr\"{o}dinger equation is different from the Liouville equation for evolution of classical distribution. This is the third point of departure from classical physics.

To summarize, we have noted three points of departure from classical physics. First, we cannot set a single point in phase space as the initial condition of the state as the state demands some spread due to the uncertainty principle. Second, even if we attempt to salvage the situation by considering the initial state as a classical probability distribution, the candidate quantum distribution turns out to be very different from a classical one, since it may not be positive definite. Third, even if we come to terms with the negative regions, the evolution equation for this distribution has higher order corrections compared to the classical Liouville equation, so that we cannot even imagine each point of the distribution as following its own classical trajectory. (These are not the only aspects of quantum behavior though. For example, operators have to be Weyl transformed before expectation values can be found by integration over phase space with the Wigner function. For operators made up of only the position operator or only the momentum operator, the Weyl transform would just replace operators by c-numbers, but the transform is non-trivial for operators containing both position and momentum \cite{Case:2008}.)

The first point above is non-negotiable and the evolution of the system can no longer be viewed as one along a definite trajectory in the phase space. However, the second point can be negotiated for certain special states. For example, Wigner functions corresponding to Gaussian states, are positive definite \cite{Curtright:2001jn}.
Moreover, the third point can also be taken care of, for the potentials which are at most quadratic in the position variable \cite{Case:2008}. So, for Gaussian Wigner functions evolving in potentials which are at most quadratic, it is only the first point that stops us from having a specific trajectory of evolution.

But there are certain unbounded potentials which have the property that an initial Gaussian Wigner function will evolve to be highly squeezed about the classical trajectory at late times. Examples are the inverted harmonic oscillator potential \cite{Guth:1985ya, Halliwell:1987eu, Albrecht:1992kf} and the free particle potential \cite{Kiefer:1998jk}.  For example, the inverted harmonic oscillator with frequency $\omega$ would lead to squeezing about the line $p-\omega x$.  But the uncertainty principle extracts its price for the high squeezing {\it about} the trajectory by demanding extreme broadening {\it along} the trajectory. Thus, in this case, although there may be large amount of squeezing  about the trajectory, the system is not completely deterministic in the sense that the probability distribution is not peaked about a particular point at a particular time, as we would expect if we try to imagine the evolution of a classical particle. Such a state is described as a ``squeezed state''.

So, at late times there is an almost infinite spread in $x$ or $p$ direction, so that a measurement of $x$ or $p$ may give any value. However, any measurement of $p-\omega x$ is expected to give zero. Thus although we do not have information of individual phase space variables, but for the inverted harmonic oscillator, distribution in phase space relates one variable to the other through an equation, which happens to be the classical equation of motion.

As noted in \cite{Albrecht:1992kf}, this is very unlike the coherent state, taken as the most classical state in areas of physics like quantum optics, and hence may even be deemed very ``non-classical''. Nevertheless, such highly squeezed states do exhibit many features that we expect from a classical system \cite{Guth:1985ya, Halliwell:1987eu, Albrecht:1992kf, Kiefer:1998jk}. The corresponding wavefunctions approach the WKB limit and the system will be practically indistinguishable from a classical stochastic system.  So in this case, instead of a deterministic classical system, what we obtain is a distribution of classical systems.

Thus, it is possible to construct systems \cite{Guth:1985ya,Kiefer:1998jk,Lombardo:1999du} which during evolution spontaneously tend to develop classical character in the above sense, without the help of any active or dynamic environmental quantum degrees of freedom.
In such systems the unboundedness of the governing potential is exploited as the system rolls down the potential so that the Wigner function gradually starts peaking on a trajectory (as a first signature of a gradually fading quantum character) {\it which is
the solution of the classical equations of motion}. Therefore, such systems acquire a classical trajectory (in the sense of probability being peaked around it in the phase space) at late times. 

There are models in the context of quantum cosmology \cite{Guth:1985ya,Singh:1989ct}, where such an emergence of the classical correlation is used as a pointer of classicality. However, whether a strong classical correlation alone is  sufficient for a system to be dubbed classical is also debated \cite{Morikawa:1990iz, Anderson:1990vc, Habib} from different points of view. 
Usually, in a typical conservative system,
there is a compromise between the spread of the Wigner function and the diagonalization of the density matrix \cite{Lombardo:1999du, Morikawa:1990iz, Laflamme:1990kd, Habib:1990hx}. In the analysis in \cite{Guth:1985ya,Singh:1989ct}, the focus is on  classical correlation without too much concern for decoherence, which makes the value of these approaches debatable. However, Kiefer et. al..\cite{Kiefer:1998pb} suggested that in the context of cosmological perturbations, analysis of the classical correlation alone should be sufficient for an apparent classicalization since the quantum coherence is expected to be suppressed anyway, owing to interaction with other fields. 

Another approach of estimating classicality of the state is to analyze its overlap with a coherent state \cite{Anderson:1990vc,Habib:1990hx}. This scheme is useful in the case of quadratic potentials but does not work for a more general case.

In this paper, we will  study whether a spontaneous classicalization can be obtained by turning on a coupling between two quantum systems which do not have unbounded Hamiltonian in the absence of the coupling.
We study the effect of interaction between two quantum systems and show that a coupling of sufficient strength between these systems can produce this dramatic outcome. For this purpose, we will first consider two simple harmonic oscillators and add a linear coupling between them. We will explore the properties of the oscillators one at a time, using reduced density matrices, reduced Wigner functions etc. constructed in a suitable manner.  Our scheme of studying the reduced functions is quite different from what is done in the context of decoherence, because our model describes an isolated system of {\it only two} harmonic oscillators, unlike decoherence which  requires a large number of environmental degrees of freedom to be integrated out if  the off-diagonal elements of the reduced density matrix are to be suppressed. Our mechanism does not rely on the existence of only a few stable pointer states which the system will be ultimately guided into. In fact, we will show that although we have an emergence of classical correlation through a sharply peaked Wigner function around the classical trajectory, the density matrix remains non-diagonal as expected for a conservative system. Because of these features this scheme is quite different --- conceptually and mathematically --- from the scheme of decoherence. 

To avoid possible misunderstandings, we stress on the fact that our main aim is to point out that an exactly solvable, model system behaves in two drastically different ways when a parameter describing the system is varied across a critical value. (We expect such a behavior to occur in more realistic systems as well, though it is difficult to demonstrate it mathematically.) We use a criterion for classicality, \textit{adapted to this specific context}, to highlight this behavioral difference in our model system. Our aim is not to come up with a criterion for classicality which is universally valid (and in fact we very much doubt whether such a criterion exists) or to debate broader philosophical aspects of different criteria for classicality which one can introduce. Instead we introduce a criterion which seems to be reasonable and highlights what we want to highlight in this specific context. This perspective provides the backdrop for the results reported in this paper and should be kept in mind.
Nevertheless, we will be able to show that the system can be considered classical, even from the perspective of other reasonable notions of classicality available in the literature. 

Our  interest, in a more general sense, is in studying the role of interactions vis-\'{a}-vis the development of effective classical features in any subsystem of interest within any arbitrary system. Since we will be interested (in future works) in generalizing this approach to a more realistic non-quadratic interaction (needed e.g., in quantum cosmology), we do not attempt a coherent state analysis here. It will be interesting to investigate the role of the back-reaction of matter fields in making some of the degrees of freedom of the universe become classical. Canonical quantization of the FRW universe \cite{DeWitt} ends up with a set-up similar to the one we are going to consider in the paper, and hence our analysis here will be a first step in that direction. Further, we argue that the number of extra degrees of freedom  is actually irrelevant for the analysis. Having shown that the classicalization can be achieved with the help of {\it just one} extra degree of freedom, we realize that this model has the potential to describe, through a more general analysis, the evolution scheme of the universe which has the bulk of its constituents becoming practically classical during later stages. Therefore, the effect of different types of coupling of the geometry with matter fields in the early epoch will be particularly interesting. That analysis is left out for a future work. 

We have organized our paper in the following form. In section II, we  set-up our  model of two harmonic oscillators with a linear coupling. We will explore its properties at the classical level and identify the interesting regimes. In section III, we study the quantum physics of this system when the coupling parameter is in one range  which we call sub-critical. In this regime, we observe the quantum behavior of the system in terms of reduced functions. We also see that the quantum domain is characterized with an emergent thermal behavior without we having traced out of a large environmental degrees of freedom. Section IV deals with the case where the coupling parameter is such that the system is super-critical. This leads to the  building up of classical correlation in one set of variables. We also test the effectiveness of this build up in terms of different  schemes. We verify that the density matrix not only loses its thermal character in this regime but also remains non-diagonal indicating the lack of a large environment. Nevertheless  the system achieves classicality. In Section V, we will discuss the connection of such a  model to more interesting physical systems and  compare and contrast information obtainable from such a model. We will also discuss the generality of our  analysis. We will summarize our results  in section VI.

\section{Model: Oscillators with linear coupling}

We will consider a system of two harmonic oscillators coupled linearly with each other and described by the Lagrangian: 
\bea
L=\frac{1}{2}m_1\dot{\tilde{x}}_1^2+\frac{1}{2}m_2\dot{\tilde{x}}_2^2-\frac{1}{2}m_1\omega_1^2\tilde{x}_1^2-
\frac{1}{2}m_2\omega_2^2\tilde{x}_2^2+\tilde{\lambda} \tilde{x}_1 \tilde{x}_2, \label{Lagrangian}
\eea
where all the parameters are real. While this is a seemingly trivial model, we will see that it has all the features we need and does lead to interesting results which are obtainable analytically. In particular, it can mimic features in more  general  models \cite{Kiefer:2009zz}, \cite{DeWitt} with unbounded Hamiltonian, for certain range of parameters.  (In fact, \cite{Kiefer:2009zz}  uses a similar model for studying the back reaction of Hawking radiation on black hole.)

With some rescaling, the above Lagrangian can be written as
\bea
L=\frac{1}{2}\dot{x}_1^2+\frac{1}{2}\dot{x}_2^2-\frac{1}{2}\omega_1^2x_1^2-
\frac{1}{2}\omega_2^2x_2^2+\lambda x_1 x_2,
\eea
where
\bea
x_{1\{2\}}=\sqrt{m_{1\{2\}}}\tilde{x}_{1\{2\}}; \hspace{0.5 in} \lambda =\frac{\tilde{\lambda}}{\sqrt{m_1m_2}}.
\eea
The momenta corresponding to $x_{1\{2\}}$ is given as 
$$p_{1\{2\}} =\dot{x}_{1\{2\}}=\frac{\tilde{p}_{1\{2\}}}{\sqrt{m_{1\{2\}}}}. $$
The classical equations of motion of this system are
\bea 
\ddot{x}_1+\omega_1^2 x_1 &=& \lambda x_2 \nonumber\\
\ddot{x}_2+\omega_2^2 x_2 &=& \lambda x_1.
\eea
We see that the two variables appear as source terms in each other's equation of motion and the coupling parameter 
governs the strength of the source term relative to the restoring force.
The Hamiltonian for this system is given as 
\bea
{\cal H}=\frac{1}{2}p_1^2+\frac{1}{2}p_2^2+\frac{1}{2}\omega_1^2x_1^2+
\frac{1}{2}\omega_2^2x_2^2-\lambda x_1 x_2. \label{CoupledHam}
\eea
The potential appearing in  \ref{CoupledHam}
\bea
V(x_1,x_2)=
\frac{1}{2}(
\begin{array}{ll}
x_1 & x_2
\end{array})
\left(
\begin{array}{ll}
\omega_1^2 & -\lambda \\ 
-\lambda & \omega_2^2
\end{array}
\right)
\left(
\begin{array}{ll}
x_1\\
x_2
\end{array}
\right)\label{Potential}
\eea
can be diagonalized by a real orthogonal transformation and written, in the diagonal basis, as
\bea
V(x_1,x_2)=V_D(X_-,X_+)=
\frac{1}{2}(
\begin{array}{ll}
X_- & X_+
\end{array})
\left(
\begin{array}{ll}
\tilde{\Omega}_-^2 & 0 \\ 
0 & \tilde{\Omega}_+^2
\end{array}
\right)
\left(
\begin{array}{ll}
X_-\\
X_+
\end{array}
\right), \label{DiagonalPot}
\eea
The two expressions for the potential are related  through the orthogonal transformation ${\cal O}$ generated by:
\bea
\left(
\begin{array}{ll}
\tilde{\Omega}_-^2 & 0 \\ 
0 & \tilde{\Omega}_+^2
\end{array}
\right)=
{\cal O}^T
\left(
\begin{array}{ll}
\omega_1^2 & -\lambda \\ 
-\lambda & \omega_2^2
\end{array}
\right)
{\cal O},
\eea
where the normal mode co-ordinates are given by 
\bea
\left(
\begin{array}{ll}
X_-\\
X_+
\end{array}
\right)=
{\cal O}^T
\left(
\begin{array}{ll}
x_1\\
x_2
\end{array}
\right),  \label{Orthogonal1}
\eea
and the explicit form of the diagonalizing, orthogonal, matrix is:
\bea
{\cal O}=
\left(
\begin{array}{ll}
c_-(\omega_1^2-\omega_2^2-\Delta) & c_+(\omega_1^2-\omega_2^2+\Delta) \\ 
-2c_-\lambda & -2c_+\lambda
\end{array}
\right)=
\left(
\begin{array}{ll}
\cos{\theta} & -\sin{\theta}\\
\sin{\theta}& \cos{\theta}
\end{array}
\right), \label{Orthogonal2}
\eea
with,
$$ c_{\pm}^2=\frac{1}{2\Delta^2\pm2(\omega_1^2-\omega_2^2)\Delta}.$$
One can verify  that the normal mode frequencies can be expressed in terms of the original frequencies and the coupling constant as 
\bea
\tilde{\Omega}_{\pm}^2=\frac{(\omega_1^2+\omega_2^2) \pm \Delta }{2},\label{eigenvalues}
\eea
with,
\bea
\Delta=\sqrt{(\omega_1^2-\omega_2^2)^2+4 \lambda^2}.
\eea
The kinetic energy part of the Hamiltonian becomes, in terms of the normal mode variables,  
\bea
\frac{1}{2}p_1^2+\frac{1}{2}p_2^2=\frac{1}{2}P_+^2+\frac{1}{2}P_-^2,
\eea
where
\bea
\left(
\begin{array}{ll}
P_-\\
P_+
\end{array}
\right)=
{\cal O}^T
\left(
\begin{array}{ll}
p_1\\
p_2
\end{array}
\right).
\eea
Thus, the Hamiltonian  can be written as that  of two
uncoupled oscillators:
\bea
{\cal H}=\frac{1}{2}P_-^2+\frac{1}{2}P_+^2+\frac{1}{2}\tilde{\Omega}_-^2X_-^2+
\frac{1}{2}\tilde{\Omega}_+^2X_+^2. \label{CoupledHam2}
\eea
This procedure is completely standard and we have essentially performed the normal mode analysis of the coupled harmonic oscillator and transformed the system to two uncoupled oscillators with   the normal mode frequencies  given by $\Omega_{\pm}$ in \ref{eigenvalues}. The original coordinates of the oscillators are related to the normal mode co-ordinates
through the orthogonal transformation \ref{Orthogonal1}. We can now study the quantum evolution of this system in a straight forward manner. Note, however, that
the normal mode frequencies remain real only as long as $\lambda < \omega_1 \omega_2$. If we parametrize $\lambda$ as 
$$\lambda=\xi \omega_1 \omega_2,$$
the criterion of the normal mode frequencies to remain real becomes $\xi<1$. 
So this problem corresponds to
that of two  harmonic ``oscillators'' only as long as $\xi<1$ (sub-critical regime). As soon as $\xi>1$ (super-critical regime), one of the normal mode frequencies, namely $\tilde{\Omega}_-$, becomes imaginary and the corresponding mode $X_-$ becomes an inverted ``oscillator'' (which, of course, does not oscillate at all!). We will discuss both these cases starting from the sub-critical coupling.

\section{Quantum Dynamics with $\xi<1$}

\subsection{Density Matrix}

As long as both the normal mode frequencies are  real, we have the standard harmonic oscillator eigenstates for the Hamiltonian in terms of $X_-,X_+$; with the quantum state of the system being  a direct product of the two individual harmonic oscillator states
$$\Psi(X_-,X_+)=\psi_1(X_+)\psi_2(X_-).$$
In particular, we can construct a Gaussian state which is the ground state of this Hamiltonian:
\bea
\Psi=\frac{1}{\sqrt{\hbar \pi}}(\tilde{\Omega}_-\tilde{\Omega}_+)^{\frac{1}{4}}\exp{\left[-\frac{(\tilde{\Omega}_-X_-^2+\tilde{\Omega}_+X_+^2)}{2\hbar} \right] }. \label{GS10}
\eea
This  state can be expressed in terms of the original variables as
\bea
\Psi=\frac{1}{\sqrt{\hbar \pi}}(\tilde{\Omega}_-\tilde{\Omega}_+)^{\frac{1}{4}}\exp{\left[-\frac{({\cal A} x_1^2+ {\cal B}x_2^2+2{\cal C}x_1x_2)}{2\hbar} \right] }, \label{GS11}
\eea
with 
\bea
{\cal A} &=& \tilde{\Omega}_- {\cal O}_{11}^2+\tilde{\Omega}_+ {\cal O}_{12}^2, \nonumber\\
{\cal B} &=& \tilde{\Omega}_- {\cal O}_{21}^2+\tilde{\Omega}_+ {\cal O}_{22}^2, \nonumber\\
{\cal C} &=& \tilde{\Omega}_- {\cal O}_{11}{\cal O}_{21}+\tilde{\Omega}_+ {\cal O}_{12}{\cal O}_{22},
\eea
using the orthogonal transformation matrix ${\cal O}$ identified in \ref{Orthogonal2}. The normalization remains the same owing to the orthogonality of the  transformation between ($X_-,X_+$) modes and ($x_1,x_2$) modes.
The corresponding ground state \textit{density matrix}  is: 
\bea 
\rho(x_1,x_1';x_2,x_2')=
\left(\frac{\tilde{\Omega}_-\tilde{\Omega}_+}{\hbar^2\pi^2}\right)^{1/2}\exp{\left[-\frac{({\cal A} x_1^2+ {\cal B}x_2^2+2{\cal C}x_1 x_2)}{2\hbar} \right] }
\exp{\left[-\frac{({\cal A}^* x_1'^2+ {\cal B}^*x_2'^2+2{\cal C}^*x_1'x_2')}{2\hbar} \right] }. \label{FullDM}
\eea
which, of course, is time-independent because  we are working  with a stationary state. We now want to  concentrate on the behavior of one of the two  modes (say $x_2$) and hence we will construct the reduced density matrix defined by:
\bea
\rho_R(x_2,x_2')=\int dx_1 \rho(x_1,x_1;x_2,x_2'),
\eea
which is obtained from \ref{FullDM} by tracing out the $x_1$ modes. On performing the integration, we get
 \bea
\rho_R(x_2,x_2')=\left(\frac{\tilde{\Omega}_-\tilde{\Omega}_+}{\hbar \pi {\cal A}_R}\right)^{1/2}
\exp{\left[-\frac{1}{2\hbar}\left(\left({\cal B}-\frac{{\cal C}^2}{2{\cal A}_R}\right) x_2^2+ \left({\cal B}^*-\frac{{\cal C}^{*2}}{2{\cal A}_R}\right) x_2'^2 
-\frac{|{\cal C}|^2}{{\cal A}_R}x_2x_2' \right)\right] }. \label{RedDM}
\eea
The parameters appearing in the reduced density matrix are all real as long as $\xi<1$. Thus the coefficients for $x_2^2$ and $x_2'^2$ are same for the sub-critical case. 
Interestingly enough, this reduced density matrix can be mapped to  a \textit{thermal} density matrix with a frequency $\omega_T$ and an inverse temperature
$\beta_T$, given by:
\bea
\rho_T(x_2,x_2')=\sqrt{\left(\frac{\omega_T}{\hbar \pi} \tan{\frac{\hbar \beta_T \omega_T}{2}} \right)}
\exp{\left[-\frac{\omega_T}{2\hbar\sin{\hbar \beta_T \omega_T} }\{(x_2^2+x_2'^2)\cos{\hbar \beta_T \omega_T} -2x_2x_2'\}\right] }, \label{ThermalDM}
\eea
provided we identify the  parameters of the thermal density matrix as
\bea
\omega_T^2 &=& \frac{{\cal B}}{{\cal A}}({\cal A}{\cal B}-{\cal C}^2), \nonumber\\
\beta_T &=& \frac{1}{\hbar \omega_T} \cosh^{-1}\left(\frac{2 {\cal A} {\cal B}}{{\cal C}^2}-1 \right).
\eea
Thus, for the mode $x_2$ we obtain a thermal character with a frequency which is different from its original frequency and a temperature which is determined by the coupling as well as the original frequencies of the system.
We can also  rewrite these quantities in terms of the parameters appearing in the original Lagrangian \ref{Lagrangian} as
\bea 
\omega_T^2 &=& \frac{\omega_1^2\omega_2^2(\omega_1^2-\omega_2^2)+\lambda^2(\omega_2^2-\omega_1^2+\sqrt{\omega_1^2\omega_2^2-\lambda^2})}
{\lambda^2+\omega_1^2(\omega_1^2-\omega_2^2)}, \nonumber\\
\beta_T\hbar \omega_T &=& \cosh^{-1}\left(\frac{2(\omega_2^2\sqrt{\omega_1^2\omega_2^2-\lambda^2}+\omega_1^2(2\omega_2^2+\sqrt{\omega_1^2\omega_2^2-\lambda^2})) -3\lambda^2}{\lambda^2}   \right). \label{ThDMPar}
\eea
Therefore, we see that when we construct the reduced density matrix starting from the ground state of the uncoupled system, i.e. the ground state of the normal modes, it becomes a thermal density matrix characterized by a temperature and a frequency which are specified in terms of the parameters of the original coupled system. That is, in the sub-critical regime the reduced density matrix acquires a thermal character for the mode $x_2$ when $x_1$ is traced out. A similar result can easily be obtained for the mode $x_1$  as well. [The thermal nature of the reduced system is also studied in \cite{HKN}, where one obtains a thermal entropy for the observable oscillator (untraced one)]. From the above expressions \ref{ThDMPar}, we can see that in the absence of coupling, i.e. in the $\lambda \rightarrow 0$ limit, the reduced density matrix becomes one with zero temperature and the frequency $\omega_2$ which  corresponds to the ground state of the untraced mode $x_2$, as to be expected. 
  
It is interesting to note that the emergent thermal behavior of the system arises due to its interaction with just a \textit{single} extra degree of freedom and we did not need to trace over a large number of degrees of freedom. 

This result is  interesting in its own right as it sidesteps the  necessity (often taken as mandatory) for a large number of environmental degrees of freedom for a system to display an apparent thermal character. This result can have interesting implications for a system in contact with a thermal bath \cite{Caldeira} and for quantum fields in curved space \cite{Padmanabhan:Cambridge}, wherein one is usually accustomed to tracing out a large number of degrees of freedom in order to obtain thermal behavior for the system of interest.  We will pursue these aspects  in a subsequent work.

\subsection{Wigner Function}
Let us next consider the Wigner function which is related to the probability distribution in phase space and defined as 
\bea
 W[X_+,X_-,P_+,P_-]=\frac{1}{\hbar}\int du dv \Psi^*[X_+-u/2,X_--v/2]e^{\left[-i\frac{P_+}{\hbar}u-i\frac{P_-}{\hbar}v\right]}\Psi[X_++u/2,X_-+v/2].
\eea
If we start with the ground state of the Hamiltonian, we get:
\bea
 W[X_-,P_-;X_+,P_+]=\frac{4}{\hbar^2} \exp{\left(-\frac{P_-^2}{\tilde{\Omega}_-\hbar}- \frac{\tilde{\Omega_-}}{\hbar}X_-^2\right)}\exp{\left(-\frac{P_+^2}{\tilde{\Omega}_+\hbar}- \frac{\tilde{\Omega_+}}{\hbar}X_+^2\right)}.\label{WFQS}
\eea

Because the state is Gaussian, the Wigner function can be taken as the probability distribution in the phase space.
We see here that the Wigner function for the ground state of the normal modes nicely separates into products of functions of canonical variables, showing complete lack of correlation between the canonically conjugate variables. This is in sharp contrast with  a classical system following a trajectory in phase space where the coordinate and momentum are correlated.
As in the previous section, it will be 
useful to look at the reduced Wigner function when we are concerned with one of the modes. To study, say $x_2$,  we construct:
\bea 
{\cal W}_R[x_2,p_2]=\int dx_1 dp_1 W[X_-,P_-;X_+,P_+],
\eea
by tracing over $x_1$. 
Using \ref{WFQS} we obtain
\bea
{\cal W}_R[x_2,p_2]=\frac{4\pi\sqrt{\tilde{\Omega}_-\tilde{\Omega}_+}\hbar}{\sqrt{(\tilde{\Omega}_-\cos^2{\theta}+\tilde{\Omega}_+\sin^2{\theta})(\tilde{\Omega}_+\cos^2{\theta}+\tilde{\Omega}_-\sin^2{\theta})}}\nonumber\\
\times
\exp{\left[-\frac{x_2^2\tilde{\Omega}_-\tilde{\Omega}_+}{\hbar(\tilde{\Omega}_-\cos^2{\theta}+\tilde{\Omega}_+\sin^2{\theta})}-\frac{p_2^2}{\hbar(\tilde{\Omega}_+\cos^2{\theta}+\tilde{\Omega}_-\sin^2{\theta})} \right]}. \label{QWigner}
\eea 

We again see  that even the reduced Wigner function remains uncorrelated in $x_2$ and $p_2$ since it  separates into a product of two functions  one depending only on $x_2$ and the other only on  $p_2$. The evolution of this system keeps the phase space variables uncorrelated
unlike for a (semi-) classical system \cite{Halliwell:1987eu,Correlation}.

To summarize, we have seen that as long as the coupling parameter is sub-critical, the system  prepared in the ground state, evolves in time in a purely quantum fashion. This quantum domain is characterized by two interesting features. First, the Wigner function shows no signature of correlation between the conjugate quantities.

Second, if only one mode is considered at a time (irrespective of the configuration of the other mode), its behavior will be that of a thermal system with a temperature and effective frequency determined by the coupling parameter value apart from the parameters of the uncoupled system at the beginning. We will next analyze the behavior of these two aspects of the system when the coupling turns super-critical.

\section{Quantum Dynamics with  $\xi>1$}

We will first build up some mathematical machinery required to describe super-critical coupling, viz. the regime $\lambda>\omega_1\omega_2$ where one of the normal mode frequency turns imaginary
and the Hamiltonian becomes unbounded from below. In this regime,  many of the notions would need to be re-examined with  interesting implications. 
We first rewrite the Hamiltonian \ref{CoupledHam2} as

\bea
{\cal H}=\frac{1}{2}P_+^2+
\frac{1}{2}\Omega_+^2X_+^2+\frac{1}{2}P_-^2-\frac{1}{2}\Omega_-^2X_-^2. \label{CoupledHam3}
\eea
with the identifications
\bea
\tilde{\Omega}_+ &\rightarrow& \Omega_+, \nonumber\\
\tilde{\Omega}_- &\rightarrow& i \Omega_-,
\eea
from \ref{CoupledHam2} to \ref{CoupledHam3}. These identifications help us to write $\Omega_-^2$ as a positive quantity in \ref{CoupledHam3}.
More specifically,
\bea
\Omega_{-}^2=\frac{\Delta-(\omega_1^2+\omega_2^2)}{2}>0. 
\eea
For this case, with $\lambda>\omega_1\omega_2$ one of the frequencies of the decoupled modes has become imaginary, so the degrees of freedom  $X_-$
has an inverted `oscillator' potential.

Even for this super-critical coupling, we can start the system in a Gaussian state for both the uncoupled oscillators in the normal coordinates.  (Such a choice of Gaussian initial state is well  motivated by the fact that ultimately we would like to proceed from the quantum mechanical description to a quantum field theoretic description wherein a Gaussian state will describe a vacuum state.) 
But unlike in the case of bounded Hamiltonian,  the  Gaussian state \ref{GS10} now fails to be an eigenstate of the system and will evolve in time. Therefore, consideration of the time evolution is needed in the super-critical case. So we consider a general initial Gaussian state of type
\bea
\Psi(X_-,X_+;0)=N \exp{\{-(\alpha X_-^2+\tilde{\alpha} X_+^2 + 2 \beta X_+ X_-)\}},\label{Initial1}
\eea
where all the parameters are real. Since the Hamiltonian becomes decoupled in $X_{\pm}$ modes, the evolution can be studied through the action of the 
propagators of these two modes. The full  propagator becomes the product of propagators for simple harmonic and inverted 
oscillators. 
Evolution of the general Gaussian state is obtained through the action of the  propagator of the system on the initial state (details in Appendix A). This time evolution, in turn, leads to a  time dependent Wigner function given by:
\bea
W=|A|^2\frac{2\pi}{\sqrt{B_RC_R-D_R^2}}\exp{\left[\frac{-2[(D_RD_I-B_IC_R)X_+-\frac{C_R P_+}{2\hbar}+\frac{D_R P_-}{2\hbar}-(D_IC_R-D_RC_I)X_-]^2}{C_R(B_RC_R-D_R^2)} \right]} \nonumber\\
\times \exp{\left[-2 \frac{X_+P_-}{\hbar} \frac{D_I}{C_R}\right]}\exp{\left[\frac{-2[\frac{P_-^2}{\hbar^2}+4C_I\frac{P_-}{\hbar}X_-+4(C_R^2+C_I^2)X_-^2]}{4C_R}\right]}\nonumber\\
\times \exp{\left[-2[(B_R + \frac{D_I^2}{C_R})X_+^2+2\frac{(D_R C_R+ D_I C_I)}{C_R}X_+X_-]\right]}, \label{GeneralWF}
\eea
with the normalization given as 
$$ A =N\frac{\sqrt{\alpha_1\alpha_2}}{\sqrt{\tilde{\alpha}k_1-\beta^2-k_1\alpha_1\beta_1}}.$$  
The parameters $B,C$ and $D$ are time dependent. Their exact expressions are given in the Appendix A. Here we will only need their asymptotic
expressions for studying  the system at late times. We now consider the two cases $\beta=0$ and $\beta\neq0$ separately.

\subsection{Wigner Function: $\beta=0$}
Let us start with the case of $\beta=0$. This  will be of particular interest if we  start the system from the ground state of the coupled system, which will be the same as  the ground states of the normal modes (before the coupling parameter is switched on).
For this case the Wigner function \ref{GeneralWF} becomes
\bea
W=|A|^2\frac{2\pi}{\sqrt{B_RC_R}}
\exp{\left[\frac{-2\left(B_IX_++\frac{P_+}{2\hbar}\right)^2}{B_R} \right]}\exp{\{-2C_RX_-^2\}} \nonumber\\
\times\exp{\{-2B_RX_+^2\}} \times \exp{\left[\frac{-1}{2C_R}\left(\frac{P_-}{\hbar}+2C_IX_-\right)^2\right]}, \label{WFS1}
\eea
with the coefficients having the asymptotic values
\bea
B_R&=&\tilde{\alpha}\frac{\frac{\Omega_+^2}{4\hbar^2}\cosec^2{\Omega_+t}}{\tilde{\alpha}^2+\frac{\Omega_+^2}{4\hbar^2}\cot^2{\Omega_+t}};
\hspace{0.5 in}
B_I=\frac{\Omega_+}{2\hbar}\frac{\left(\frac{\Omega_+^2}{4\hbar^2}-
\tilde{\alpha}^2\right)\cot{\Omega_+t}}{\tilde{\alpha}^2+\frac{\Omega_+^2}{4\hbar^2}\cot^2{\Omega_+t}}, \nonumber\\
C_R&=&\alpha\frac{\frac{\Omega_-^2}{4\hbar^2}\cosech^2{\Omega_-t}}{\alpha^2+\frac{\Omega_-^2}{4\hbar^2}\coth^2{\Omega_-t}}\hspace{0.1 in}
\xrightarrow{\Omega_-t\gg1}\hspace{0.1 in} \alpha\frac{\Omega_-^2}{4\hbar^2}\frac{4e^{-2\Omega_-t}}{\alpha^2+\frac{\Omega_-^2}{4\hbar^2}},
\nonumber\\
C_I&=&-\frac{\Omega_-}{2\hbar}\frac{\coth{\Omega_-t}\left(\frac{\Omega_-^2}{4\hbar^2}+
\alpha^2\right)}{\alpha^2+\frac{\Omega_-^2}{4\hbar^2}\coth^2{\Omega_-t}}\hspace{0.1 in}\xrightarrow{\Omega_-t\gg1}\hspace{0.1 in} -\frac{\Omega_-}{2\hbar}. \label{Parameters2}
\eea
These expressions determine the profile of the system in the phase space as time evolves.  We will now focus on the two constituent modes separately.

\subsubsection{Reduced Wigner function: Mode $x_2$}
To study the mode $x_2$,  we translate everything back in  to the coupled basis and obtain the reduced Wigner function for the $x_2$ mode by integrating the full Wigner function \ref{WFS1} over the $x_1,p_1$ coordinates of the phase space. This calculation yields (in units with $\hbar=1$),   in the asymptotic limit of $\Omega_-t\gg1$ (when $C_R\rightarrow 0$) the result
\bea
{\cal W}_R(x_2,p_2)\sim \exp{\left[-\frac{2B_R(p_2+2C_Ix_2)^2\sec^2{\theta}}{4B_I^2+4B_R^2-8B_IC_I+4C_I^2} \right]}. \label{RWF1}
\eea
This describes the dominant behavior of the reduced Wigner function at late times. (To be precise, there is another term in the argument of exponential which is quadratic in $x_2$, whose strength gradually diminishes with $C_R\rightarrow 0$ during the course of evolution. The phase space distribution remains well-defined due to a corresponding compensation in the normalization; see Appendix A). Here, we will concentrate on the properties of the dominant part at late times,  which  captures the possible emergence of  classical correlations.
The Wigner function, in fact,  peaks at the ``classical trajectory'' if the variance 
\bea
\sigma^2=\frac{(B_I-C_I)^2+B_R^2}{B_R}\cos^2{\theta}, \label{Variance}
\eea
becomes small in some appropriate sense. We will now make this idea quantitative.
For this purpose, we can study the behavior of the reduced Wigner function by comparing this variance in \ref{Variance} to that of a typical quantum state of $x_2$ mode. We want the variance in the trajectory to be small so that the Wigner function becomes proportional to a Dirac delta distribution around the classical trajectory (see Appendix D). Moreover, we wish to obtain a criterion for classicality in a dimensionless form. 

For a system with a deterministic, classical, trajectory, we can find a value of $p$ for a given value of $x$. But for a quantum system described by a Wigner function, there will be a spread in the possible values of $p$ at any given $x$. If this spread is smaller than a typical spread we expect in the quantum state of the system, we can consider it a signal for build up of classical correlation in phase space. 
More precisely, for every  point on the trajectory (with a given $x$), we can calculate the variance in the $p$-value, which can be  compared with the typical variance in $p$ for a quantum system. The ground state of any system will be a natural choice for a typical quantum state. Thus, we can take a typical quantum state for $x_2$ mode to have a variance in $p_2$ of about
$$\sigma_Q^2\sim \omega_2 \hbar.$$
We can see from \ref{Parameters2} and \ref{Variance} that the variance in the Gaussian for the reduced Wigner function is an oscillatory function varying between a minimum and maximum values at late times.
We evaluate the maxima of the variance and demand it to be small when compared to that of a typical quantum state. 
We can see that as long as $\tilde{\alpha}>\Omega_+/2\hbar$, the variance gets maximized for $\cot{\Omega_+t}=-\Omega_+/\Omega_-$ at late times, with the maximum value ( $\hbar=1$ ):
\bea
\sigma^2_M=\tilde{\alpha}\left(1+\frac{\Omega_-^2}{\Omega_+^2}\right)\cos^2{\theta}.
\eea
Therefore, our criterion for classical correlation  gets translated into the condition:
\bea
\frac{\tilde{\alpha}(\Omega_-^2+\Omega_+^2)}{\Omega_+^2 \omega_2}\cos^2{\theta}\ll1. \label{CCR01}
\eea
Using the previously defined expressions for $\theta$ and $\Omega_\pm$ \ref{Orthogonal2}, \ref{eigenvalues} and writing
\bea
\lambda=\xi\omega_1\omega_2  \hspace{0.25 in} \text{and} \hspace{0.25 in} \omega_1=f \omega_2,
\eea
with $\xi>1$, the condition \ref{CCR01} can be expressed as,
\bea
\frac{\omega_2}{\tilde{\alpha}}\gg\frac{(\sqrt{(f^2-1)^2+4\xi^2f^2}-(f^2-1))}{\sqrt{(f^2-1)^2+4\xi^2f^2}+(f^2+1)}.\label{CCR11}
\eea
Thus, the condition for emergence of classical correlation in the Wigner function relates
the dispersion of the mode $x_2$ to that of one of the uncoupled modes $X_+$. 
The variance for the regime $\tilde{\alpha}<\Omega_+/2\hbar $, attains the maximum value whenever  $\cot{\Omega_+t}=\Omega_-/\Omega_+$, with the maximum value
\bea
\sigma^2_M=\left(\frac{\Omega_+^2+\Omega_-^2}{4\tilde{\alpha}}\right)\cos^2{\theta}.
\eea
In this regime, the criterion of strong classical correlation will become
\bea
\frac{\tilde{\alpha}}{\omega_2}\gg \frac{\sqrt{(f^2-1)^2+4\xi^2f^2}-(f^2-1)}{8}. \label{CCR12}
\eea
Thus we have identified the region of the parameter space which will make an initial Gaussian state evolve to a sharply peaked Wigner function. If we decide to prepare the initial state  in the state analogous to the ground state, this will correspond to the choice,
$$\tilde{\alpha}=\frac{\Omega_+}{2\hbar}.$$
In this case we can verify the variance does not oscillate at late times and hence its value can be used directly in our criterion. One can also verify, through some straightforward analysis, that in this case,
$$\frac{\sigma^2_M}{\omega_2}=\frac{[\sqrt{(f^2-1)^2+4\xi^2f^2}-(f^2-1)]}{2\sqrt{2}[\sqrt{(f^2-1)^2+4\xi^2f^2}+(f^2+1)]^{1/2}},$$
so that the limit $f\gg\xi^2>1$ satisfies the criterion for the existence of classical correlation.
It is important to note that we get an emergent correlation between the phase space variables in the reduced Wigner function when the system parameters satisfy some condition.
Further, the larger is the separation between $\omega_1$ and $\omega_2,$ (i.e. the larger is the value of
$f$) the better is the correlation. \textit{However,  although for large $f$ the variance gets smaller, it remains a non-zero constant, as long as $f$ is finite, irrespective of how long we wait.} This suggests that the mode $x_2$ does not get ``more classical" than a particular limit. There is always some amount of residual (quantum) variance left. 
We know from the transformation \ref{Orthogonal1}, that the modes $x_1$ and $x_2$ can be thought of as been made up by combining the normal modes $X_{\pm}$. We see that the interesting case, viz. the one in which classical correlation emerges, involves aligning $x_2$ along $X_-$ more and more effectively, which (as a normal mode) attains vanishing variance  at very late times.

\subsubsection{Reduced Wigner function: Mode $x_1$}

Let us next study the fate of the other mode $x_1$ under the evolution and test its behavior from the point of view of the Wigner function. We can again do the same analysis as before, by  tracing over $x_2, p_2$ coordinates of the phase space. Starting with \ref{WFS1} we do the integration over $x_2,p_2$ to obtain, in the late time limit (which coincides with $C_R\rightarrow 0$, $D_R\rightarrow 0$) the result:
\bea
{\cal W}_R(x_1,p_1)\sim \exp{\left[-\frac{2B_R(p_1+2C_Ix_1)^2\cosec^2{\theta}}{4B_I^2+4B_R^2-8B_IC_I+4C_I^2} \right]}. \label{WFS2}
\eea 
As before, we do see a possibility of the build up of the correlation with a variance which is oscillatory at late times.  Just as in the case of $x_2$, we again have two different regimes $2\tilde{\alpha}< \Omega_+$ and $2\tilde{\alpha}> \Omega_+ ,$ where the variance of the trajectory for $x_1$ modes attains two different maxima values. We can again obtain the expression for $\sigma^2_M/\omega_1$ (in which we compare the variance to that in a typical quantum state corresponding to $x_1$) as
\bea
\frac{\sigma^2_M}{\omega_1}=\frac{\tilde{\alpha}[\sqrt{(f^2-1)^2+4\xi^2f^2}+(f^2-1)]}{\omega_2 f[\sqrt{(f^2-1)^2+4\xi^2f^2}+(f^2+1)]}
 \text{\hspace{0.25 in}for \hspace{0.25 in}} 2\tilde{\alpha}> \Omega_+,
\eea
which, for a strong correlation, needs to satisfy
\bea 
\frac{\omega_2}{\tilde{\alpha}} \gg \frac{[\sqrt{(f^2-1)^2+4\xi^2f^2}+(f^2-1)]}{ f[\sqrt{(f^2-1)^2+4\xi^2f^2}+(f^2+1)]}. \label{CCR31}
\eea
Similarly in the regime $2\tilde{\alpha}< \Omega_+ $ we have
\bea
\frac{\sigma^2_M}{\omega_1}=\frac{[\sqrt{(f^2-1)^2+4\xi^2f^2}+(f^2-1)]\omega_2}{8f \tilde{\alpha}} ,
\eea
 which, for a strong correlation will require 
\bea 
\frac{\tilde{\alpha}}{\omega_2} \gg \frac{[\sqrt{(f^2-1)^2+4\xi^2f^2}+(f^2-1)]}{8f}. \label{CCR32}
\eea
For both  $x_1$ and $x_2$ modes to simultaneously develop classical correlations  we need to satisfy (i) both \ref{CCR11} and \ref{CCR31} in the regime $2\tilde{\alpha}> \Omega_+$ or (ii) \ref{CCR12} and \ref{CCR32} in the regime $2\tilde{\alpha}< \Omega_+ $. Clearly, this demands a very special initial configuration in terms of values of $\tilde{\alpha}$ and hence simultaneous classicality for both the modes  will not be feasible for a typical state. Analysis done for the constant variance case at $2\tilde{\alpha}=\Omega_+$, when we start with a ground state in the normal modes, shows that 
$$\frac{\sigma^2_M}{\omega_1}=\frac{\sqrt{2}[\sqrt{(f^2-1)^2+4\xi^2f^2}+(f^2-1)]}{4f[\sqrt{(f^2-1)^2+4\xi^2f^2}+(f^2+1)]^{1/2}} $$
which in the limit $f\gg\xi^2>1$ (which induces the correlations between $x_2$ and $p_2$) turns out to be of order unity,
\bea 
\frac{\sigma^2_M}{\omega_1}=\frac{\sqrt{(f^2-1)^2+4\xi^2f^2}+(f^2-1)}{2\sqrt{2} f[\sqrt{(f^2-1)^2+4\xi^2f^2}+(f^2+1)]^{1/2}}\sim {\cal O}(1).
\eea

This shows that the correlation  between $x_1$ and $p_1$ has not developed sufficiently in this limit and the region of parameter values we identified for classicality of $x_2$ mode is \textit{ not} compatible with the corresponding region of interest  for $x_1$. So, while the mode $x_2$ starts behaving classically in appropriate limit, the mode $x_1$ retains the quantum nature since its correlation build up is feeble. One can, in principle, debate our choice of selecting a particular dimensionless parameter to define classicality; but we can clearly see that the relative variance in $x_1$ mode is much larger than that of $x_2$. As discussed above, the limit $f\gg\xi^2>1$ `aligns' the variables $(x_1,x_2)$  towards $(X_+,X_-)$ and hence the mode $x_1$ is expected to remain non-classical.

We thus  see that when we study the reduced Wigner functions for the modes $x_1$ and $x_2$, the Wigner function in one of the modes becomes strongly peaked around a trajectory which satisfies the classical equations of motion, while there is no significant peaking for the reduced Wigner function of the other mode. Such a trend is reversed if we take $\xi^2\gg f>1$. This fact indicates that one of the modes (with the lower frequency) turns classical, while the other (large frequency) mode remains  quantum mechanical. Although we have chosen the sharpness of Wigner function as the signature of classicality for the mode $x_2$, this outcome can also be motivated through some other techniques of gauging the classicalization in such a system. We will briefly discuss two such prominent proposals in literature.

\textit{1. Proposal of Morikawa:}

This proposal \cite{Morikawa:1990iz} studies the classicality of a system through two criteria, Classical Correlation (CC) and Quantum Decoherence (QD). In a Wigner function of the type
\bea
W(x,p)\sim\exp{\left[-\frac{(p-\bar{\beta} x)^2}{4 \bar{\gamma}^2}-\bar{\alpha}^2 x^2 \right]}, \label{Morikawa}
\eea
the peaking of the Classical Correlation (CC) will be occurring if 
\bea
\delta_{CC}\sim\frac{\bar{\alpha} \bar{\gamma}}{\bar{\beta}}\ll1,\label{CC}
\eea
which, in a sense, measures the variance in the variables with respect to their expectation values.
Further, for a system to be characterized as classical, it is  required that a separate criterion of Quantum Decoherence (QD)
\bea
\delta_{QD}\equiv\frac{\bar{\alpha}}{\bar{\gamma}} \ll1,\label{QD}
\eea
is also satisfied. This parameter
measures the ratio of quantum coherence length in the system to the typical size of the system. For a system to qualify as being classical, Morikawa suggested that both  the criteria CC and QD have to be satisfied simultaneously. This demand is not satisfied in typical conservative systems owing to their apparently incompatible structure, viz. CC increases with $\bar{\gamma}$ while QD decreases with $\bar{\gamma}$.

In our case we see that reduced Wigner function of $x_2$ assumes the form \ref{RWF1} asymptotically. We can see that while the parameters analogous to $\bar{\beta}$ and $\bar{\gamma}$ approach a non-zero constant value asymptotically, the parameter analogous to $\bar{\alpha}^2$ gets vanishingly small in the asymptotic limit. This makes the case for the original mode $x_2$ quite interesting as both  criteria of classicality get satisfied in the late time limit. After a sufficiently long  but finite time the modes $x_2$ becomes effectively classical in this sense.

\textit{2. Classicality parameter:}

Following the discussion in \cite{Mahajan:2007qc}, we can associate another parameter to measure classicality with our reduced Wigner functions. The 
idea behind this scheme is to quantify the strength of correlation between the phase space variables. Formally, it is defined as 
\bea
{\cal C}=\frac{\langle p q \rangle_W}{\sqrt{\langle p^2 \rangle_W\langle q^2 \rangle_W}},
\eea
where the symbol $\langle \hspace{0.1 in} \rangle_W$ represents the averaging over the Wigner function.
The limit ${\cal C}\rightarrow 1$ marks a strong correlation between the canonical variables and signals classicality, whereas the other limit ${\cal C}\rightarrow 0$ suggests that there is effectively no  correlation. If we use this parameter for classicality criterion, we obtain for a Wigner function of kind \ref{Morikawa},
\bea
{\cal C}=\frac{\bar{\beta}}{\sqrt{(4\bar{\gamma}^2 \bar{\alpha}^2+\bar{\beta}^2)}}.
\eea

We once again see here that this parameter
vanishes for \ref{QWigner} since the value of parameter $\bar{\beta}$ vanishes. Thus as long as the coupling parameter is below the threshold, this parameter also suggests that there is no notion of classical correlation. 
On the other hand, we can see from \ref{RWF1} that in the late time limit $\bar{\alpha}\rightarrow 0$ (for more details refer to Appendix A), we obtain ${\cal C} \rightarrow 1$, which suggests a gradual building up of a strong classical correlation.

\textit{3. Discussion:}

It should, however, be stressed  that these two proposals  are insensitive to the value of $\bar{\gamma}$. So they suggest that, irrespective of the value of $f$, there is always an emergent classical correlation (classicality) if we wait long enough and if the coupling strength is super-critical. The limitations of the  two proposals become further apparent if we scrutinize the behavior of mode $x_1$ according to them. We immediately see that since the reduced Wigner function has a functional behavior qualitatively the same as before in the late time limit ($C_R \rightarrow 0$). The criteria set by these two schemes will be satisfied, suggesting that even the mode $x_1$ develops a classical character (with a time scale different from the time scale for the mode $x_2$ in both the schemes). But we have already seen that the corresponding reduced Wigner function does not peak around a trajectory in the phase space leading to the absence of any appreciable correlation between the corresponding conjugate variables. But the two criteria discussed above suggest a complete classicalization of the full system at late times. Since this conclusion challenges the notion of a specific trajectory as a prominent feature of classicality, we prefer to stick with  our choice (of considering the 
variance around the classical trajectory) as a measure of classicalization for the rest of the paper.

These results show that our model indeed shows classical character (at least partially, if not fully for the system) at late times, irrespective of the criterion used to test the classicality the system.
If we start with the initial state prepared in a Gaussian form in the normal modes, with the width of the Gaussian in $X_{+}$
similar to the one determined by the normal mode frequency, there exists a parameter range in which one of the coupled modes acquires classicality, while the other mode remains as quantum mechanical. Furthermore, the degree of classicality for this mode (identified with the one with lower frequency) is not achieved to arbitrary precision; certain degree of ``quantumness" (measured by the residual non-zero variance around the classical trajectory) remains. This ``quantumness''  is also a measure of the coupling between the  coupled modes. As the strength of the coupling is increased, keeping the frequencies constant, we move from this regime to the one in which quantum features of $x_2$ get enhanced. All these go to emphasize a very interesting role of the coupling. No classicality emerges until the coupling strength increases above a certain critical value. When the coupling parameter becomes super-critical, one of the normal modes ``tips over" and we see a spontaneous emergence of classicality in the Wigner function. For a range of value of parameters of the theory,  this peaking of the Wigner function becomes very effective. Moreover, the coupling ensures that  some amount of 
quantum nature is retained by the mode which turns classical even at arbitrarily later times. 

\subsection{Wigner function: $\beta\neq 0$}

This case  deals with a more general scenario and includes, for instance, the situation in which  the initial state is the ground state  of the original modes (rather than the normal modes). Similar analysis now leads to an expression of the Wigner function as in \ref{GeneralWF}, with the late time behavior of the parameters given by:
\bea
B&\rightarrow& -\left[\frac{\frac{i\Omega_+}{2\hbar}\tilde{\alpha}\cot{\Omega_+t}-\frac{\Omega_+^2}{4\hbar^2}
-\frac{\beta^2}{\alpha^2+\frac{\Omega_-^2}{4\hbar^2}}\left(\alpha+\frac{i\Omega_-}{2\hbar}\right)
\frac{i\Omega_+}{2\hbar}\cot{\Omega_+t}}{(\tilde{\alpha}-\frac{i\Omega_+}{2\hbar}\cot{\Omega_+t})-
\frac{\beta^2}{\alpha^2+\frac{\Omega_-^2}{4\hbar^2}}\left(\alpha+\frac{i\Omega_-}{2\hbar}\right)}\right], \label{BValues}
\\
D&\rightarrow& -\left[\frac{\beta \frac{\Omega_+\Omega_-}{2 \hbar^2}\exp{(-\Omega_-t)}\cosec{\Omega_+t}}
{\alpha\tilde{\alpha}-\beta^2-\frac{\Omega_+\Omega_-}{4 \hbar^2}\cot{\Omega_+t}-i(\frac{\Omega_-\tilde{\alpha}}{2\hbar}+
\frac{\Omega_+\alpha}{2\hbar}\cot{\Omega_+t})}\right] \rightarrow 0,  \label{DValues}
\\
C&\rightarrow& -\left[\frac{\frac{\alpha\Omega_-^2 \exp{(-2\Omega_-t)}}{\hbar^2}}{\alpha^2+\frac{\Omega_-^2}{4\hbar^2}}+i\frac{\Omega_-}{2\hbar}\right]\rightarrow
-i\frac{\Omega_-}{2\hbar}. \label{CValues}
\eea
As a consistency check we can verify that $\beta=0 \Rightarrow D=0,$ reproduces standard  results for the uncoupled case and  that the expressions remain finite at all times. 
With the late time behavior of the parameters appearing in the Wigner function encoded in the relations \ref{BValues}, \ref{DValues} and \ref{CValues} we  look at the late time reduced Wigner function which is again of the form \ref{WFS1}. The evolution drives the system to the uncoupled limit $D\rightarrow0$, although the parameters $B_R,B_I$ in this case behave somewhat differently. We can  immediately see  that the expression for the variance  remains the same as \ref{Variance}.  One can  verify that $B_R$ does not ever vanish, $B_I$  remains finite during oscillation and  the variance  is again an oscillatory function which has a finite maximum value. Therefore, as before, the dominant late time behavior of the reduced Wigner function is one with a correlation between canonical variables with the strength of the correlation oscillating between a maximum and a minimum value.  The extrema of the variance for $x_2$ have the values
\bea
\sigma_M^2=
\frac{\left(\text{$\Omega_-$}^2+\text{$\Omega_+$}^2\right)}{2
   \text{$\Omega_+$}^2 \left(\text{$\tilde{\alpha}$}-\alpha  \text{$\tilde{\beta}$}^2\right)} \left(\left(\text{$\tilde{\alpha}
   $}-\alpha  \text{$\tilde{\beta}$}^2\right)^2+\text{$\tilde{\beta}$}^4 \frac{\text{$\Omega_-$}^2}{4\hbar^2}+\frac{\text{$\Omega_+$}^2}{4\hbar^2} \pm \tilde{\Delta}\right), \label{GenVar}
\eea
with,
\bea 
\tilde{\Delta}=\sqrt{\left(\left(\text{$\tilde{\alpha} $}-\alpha  \text{$\tilde{\beta}$}^2\right)^2+\text{$\tilde{\beta}$}^4 \frac{\text{$\Omega_-$}^2}{4\hbar^2}\right)^2+\frac{\text{$\Omega_+$}^4}{16\hbar^4}-2 \frac{\text{$\Omega_+$}^2}{4\hbar^2}
\left(\left(\text{$\tilde{\alpha}$}-\alpha  \text{$\tilde{\beta}$}^2\right)^2-\text{$\tilde{\beta}$}^4\frac{\text{$\Omega_-$}^2}{4\hbar^2}\right)},
\eea
and 
$$\tilde{\beta}^2=\frac{\beta^2}{\alpha^2+\frac{\Omega_-^2}{4\hbar^2}}.$$
We will now consider some specific cases.

\subsubsection{Ground state of the coupled modes in the absence of the coupling}

If we take the initial state as the ground state of the Hamiltonian in the coupled modes in the absence of the coupling, i.e.
\bea
\Psi_0(x_1,x_2)=N\exp{\left(-\frac{\omega_1}{2\hbar}x_1^2 -\frac{\omega_2}{2\hbar}x_2^2\right)}, \label{GS02}
\eea
it can be easily verified that
\bea 
\alpha &=& \frac{\omega_1}{2\hbar}\cos^2{\theta}+\frac{\omega_2}{2\hbar}\sin^2{\theta}, \\
\tilde{\alpha}&=&\frac{\omega_1}{2\hbar}\sin^2{\theta}+\frac{\omega_2}{2\hbar}\cos^2{\theta}, \\
\beta&=&\frac{(\omega_2-\omega_1)}{2\hbar}\cos{\theta}\sin{\theta}.
\eea

It is not difficult to show that even with this initial state the variance of the trajectory for $x_2$ remains extremely small in the regime $f\gg\xi^2>1$, whereas the variance in the trajectory for the mode $x_1$ remains large.
Therefore, the emergence of the classical correlation in one of the modes appears to be a more general feature not specific to the initial state we chose previously. \footnote{The approaches of \cite{Morikawa:1990iz} and \cite{Mahajan:2007qc} also suggest the classicalization of the  system even in this case, owing to the structure of the Wigner function which is still of the form \ref{RWF1}.}.

\subsubsection{Back to sub-criticality}

We next study the state in \ref{GS02} when the coupling parameter is sub-critical, i.e. $\xi<1$.  In the subcritical case the straight line fails to be the classical equation of the trajectory for $x_2$. Moreover, there is no meaningful ``reduced description'' for either of the modes as there is no regime
where the contribution of coupling becomes insignificant. We show that the reduced Wigner function does not show any efficient peaking about the trajectory (of the super-critical case), as expected.

The general Wigner function will again be of the form of \ref{GeneralWF}, with the expressions for parameters $B,C,D$ having oscillatory behavior
\bea
B &=&-\left[\frac{i \Omega_+}{2\hbar}\cot{\Omega_+t}-\frac{(\alpha-\frac{i\Omega_-}{2\hbar}\cot{\Omega_-t})\frac{\Omega_+^2}{4\hbar^2}\cosec{\Omega_+t}}
{\alpha\tilde{\alpha}-\beta^2-\frac{\Omega_+\Omega_-}{4 \hbar^2}\cot{\Omega_-t}\cot{\Omega_+t}-i(\frac{\Omega_-\tilde{\alpha}}{2\hbar}\cot{\Omega_-t}+
\frac{\Omega_+\alpha}{2\hbar}\cot{\Omega_+t})} \right]\label{Param2.1}\\
D &=& -\left[\frac{\beta \frac{\Omega_+\Omega_-}{4 \hbar^2}\cosec{\Omega_-t}\cosec{\Omega_+t} }{\alpha\tilde{\alpha}-\beta^2-\frac{\Omega_+\Omega_-}{4 \hbar^2}\cot{\Omega_-t}\cot{\Omega_+t}-i(\frac{\Omega_-\tilde{\alpha}}{2\hbar}\cot{\Omega_-t}+
\frac{\Omega_+\alpha}{2\hbar}\cot{\Omega_+t})}\right],\label{Param2.2}\\
C &=& -\left[\frac{i\frac{\Omega_-\alpha}{2\hbar}\cot{\Omega_-t}- \frac{\Omega_-^2}{4 \hbar^2}}{\alpha-i\frac{\Omega_-}{2\hbar}\cot{\Omega_-t}}+
\frac{\beta^2\alpha_2^2}{k_1(\tilde{\alpha}k_1-\beta^2-k_1\alpha_1\beta_1)}\right], \label{Param2.3}
\eea
where we have
\bea
\alpha_1&=&\frac{i\Omega_+}{2\hbar\sin{\Omega_+t}}; \hspace{0.55 in} \beta_1=\cos{\Omega_+t} \nonumber\\
\alpha_2&=&\frac{i\Omega_-}{2\hbar\sin{\Omega_-t}}; \hspace{0.55 in} \beta_2=\cos{\Omega_-t}\nonumber\\
k_1&=&\alpha-\alpha_2\beta_2.
\eea
As before, it will be instructive to study the case for $\beta=0$ first, which is not qualitatively very different from $\beta \neq 0$ whose quantitative details we provide in the Appendix B. One can show that the Wigner function for this system, when reduced in favor of $x_1$ mode, has a form similar to \ref{Morikawa} with
\bea
\bar{\gamma}^2 &=&\frac{(B_R\cos^2{\theta}+C_R\sin^2{\theta})}{2\{(B_R-C_R)^2+(B_I-C_I)^2\}\sin^2{\theta}\cos^2{\theta}-2B_RC_R},\nonumber\\
\bar{\alpha}^2&=&\frac{2B_RC_R\left[\{(B_R-C_R)^2+(B_I-C_I)^2\}\sin^2{2\theta}+4B_RC_R\right]}{[2\{(B_R-C_R)^2+(B_I-C_I)^2\}\sin^2{\theta}\cos^2{\theta}-2B_RC_R](B_R\cos^2{\theta}+C_R\sin^2{\theta})},\nonumber \\
\bar{\beta} &=& \frac{2(B_RC_I\cos^2{\theta}+C_R B_I\sin^2{\theta})}{(B_R\cos^2{\theta}+C_R\sin^2{\theta})},
\eea
and for mode $x_2,$ we have:
\bea
\bar{\gamma}^2 &=&\frac{(B_R\sin^2{\theta}+C_R\cos^2{\theta})}{2\{(B_R-C_R)^2+(B_I-C_I)^2\}\sin^2{\theta}\cos^2{\theta}-2B_RC_R}, 
\eea
with the same $\bar{\alpha}$ and $\bar{\beta}$ as before. We see from \ref{Param2.1}, \ref{Param2.2} and \ref{Param2.3} that all the parameters remain oscillatory instead of some of them vanishing at asymptotic times. We also note that the variance term for both the modes contain $\sin{\theta}$ as well as $\cos{\theta}$ terms in the numerator, unlike in the super-critical case, which keeps the variance at non-vanishing values due to their linear independence.  With these expressions we can see that the variance of the distribution about the straight line in the phase space does not get  vanishingly small as in the super-critical case. The alternative criteria of classicality (of \cite{Morikawa:1990iz} and \cite{Mahajan:2007qc}) also leads to the conclusion that the  system remains non-classical in the sub-critical case. Therefore, we see that although the functional behavior of the reduced Wigner function remains of the type in 
\ref{RWF1}, suggestive of a peaking about the trajectory (of the super-critical case), the efficiency of peaking remains poor in the sub-critical case.

\subsection{Density Matrix}

In this subsection we discuss the fate of the thermal character of the density matrix once the coupling has crossed the threshold. For a state of the kind 
\bea
\Psi=A\exp{\{-(B X_+^2+C X_-^2 + 2 D X_+X_-)\}}, \label{GnGauss}
\eea
 the density matrix is given by
\bea
\rho=|A|^2\exp{\{-(B^* X_+^{'2}+C^* X_-^{'2} + 2 D^* X_+'X_-')-(B X_+^2+C X_-^2 + 2 D X_+X_-)\}}.
\eea
 We have dropped the arguments in the left hand sides of the above equations just for convenience.
 Using the transformation between $\{X_+,X_-\}$ and $\{x_1,x_2\}$  modes in \ref{Orthogonal2} and integrating the trace of the density matrix, over $x_1$ modes, we obtain the reduced density matrix in favor of $x_2$ mode as
\bea
\rho&\sim& \exp{\left[-\left(B^*\cos^2{\theta}+C^*\sin^2{\theta}+2D^*\sin{\theta}\cos{\theta}-
\frac{[(C^*-B^*)\sin{\theta}\cos{\theta}+D^*\cos{2\theta}]^2}{2B_R\sin^2{\theta}+2C_R\cos^2{\theta}-4D_R\sin{\theta}\cos{\theta} }\right)x_2^2\right]}\nonumber\\
&\times&\exp{\left[-\left(B\cos^2{\theta}+C\sin^2{\theta}+2D\sin{\theta}\cos{\theta}-
\frac{[(C-B)\sin{\theta}\cos{\theta}+D\cos{2\theta}]^2}{2B_R\sin^2{\theta}+2C_R\cos^2{\theta}-4D_R\sin{\theta}\cos{\theta} }\right)x_2^{'2}\right]}\nonumber\\
&\times&\exp{\left[\frac{2|(C^*-B^*)\sin{\theta}\cos{\theta}+D^*\cos{2\theta}|^2}{2B_R\sin^2{\theta}+2C_R\cos^2{\theta}-4D_R\sin{\theta}\cos{\theta}}x_2x_2'\right]}.
\label{DensityM}
\eea
For \ref{DensityM} to resemble that of a thermal density matrix,  we first require the coefficients (in the exponent) of $x_2^2$ and $x_2^{'2} $ to be equal. We observe that the coefficient of $2x_1x_2$ is already real, so if the coefficients of  $x_2^2$ and $x_2^{'2}$ become identical, then one can bring the reduced density matrix in the form of \ref{RedDM} which also has two coefficients. This criterion can be rewritten as
\bea
B\cos^2{\theta}+C\sin^2{\theta}+2D\sin{\theta}\cos{\theta}-
\frac{[(C-B)\sin{\theta}\cos{\theta}+D\cos{2\theta}]^2}{2B_R\sin^2{\theta}+2C_R\cos^2{\theta}-4D_R\sin{\theta}\cos{\theta} }\in\mathbb{R}, \label{ThermalityCondition}
\eea
which --- in the late time limit i.e. $C_R\rightarrow 0$, $D\rightarrow 0$ --- gives
\bea
(B_R+iB_I)\cos^2{\theta}+iC_I\sin^2{\theta}-\frac{[(iC_I-B_R-iB_I)\sin{\theta}\cos{\theta}]^2}{2B_R\sin^2{\theta}}\in\mathbb{R},
\eea
yielding the condition
\bea
C_I\rightarrow 0.\label{CCvsTh}
\eea

This criterion is \textit{not} satisfied in the most general case. It is a special case of $\xi=1$ where the coupling just turns critical. We note that the criterion of the thermality of the reduced density matrix \ref{ThermalityCondition} relates the parameters in the
wavefunction to the parameters of the theory $(f,\xi)$.
For the interesting range $f\gg\xi^2>1$ the criterion \ref{ThermalityCondition} gets reduced to 
\bea
B_R C_I = D_R D_I,
\eea
for a general Gaussian with $B_R \neq 0$. In the super-critical limit the coefficient $D\rightarrow 0$ generically, and thus we obtain the criterion \ref{CCvsTh} for a general Gaussian state. In the sub-critical case when we have the normal mode ground state as the initial state, the above  criterion for thermality is readily satisfied, as discussed in section III. Therefore, we see that onset of super-criticality is marked with the disappearance of the thermal character of the reduced density matrix for the individual modes.

It is also important to note that $C_I$  is essentially the source of the correlation \ref{RWF1}. Any non-zero
$C_I$  triggers correlation on the one hand, while destroying the possible thermal character of the density matrix at late times on the other hand. 
In other words, the thermal character of the reduced density matrix is lost along with the emergence of spontaneous classical correlation for $x_2$. If we do the analysis for the reduced density matrix of mode $x_1$ we again get the same condition as in \ref{CCvsTh}. Therefore, onset of classicality in the modes  destroys the thermal character in both the modes. This seems
to be a strong indicator of correlation, because note that we  started with a general Gaussian initial state \ref{GnGauss}, whose reduced density matrix would have turned thermal in appropriate limit, if $C_I\rightarrow 0$, i.e. $\bar{\beta} \rightarrow 0$ which generates correlations in the first place. That is to say, if no correlation is generated, we can always have a reduced density matrix which is thermal.

As discussed before, the emergence of classical correlation characterized  by this approach is not necessarily the decoherence limit because of the lack of environment. It can be verified from the expression \ref{DensityM} that the off-diagonal terms are not negligible in the regime $f\gg\xi^2>1$, as would have been in the case of decoherence. The classical behavior of mode $x_2$ in our case is linked only with the   emergence of classical correlations. This approach suggests that only one of the two modes remains quantum throughout in the absence of development of any effective correlation. (For more detailed discussion of this approach see \cite{Singh:1989ct} and references therein.)

Let us summarize the discussion so far.
We have established that the  emergence of classical character occurs spontaneously when the coupling constant turns critical. The parameters in the theory seem to be effective for deciding the strength of classical correlation irrespective of  whether the initial state is a ground state of either the coupled modes or the uncoupled modes. In fact, as we will discuss later in section V, the emergence of the classical correlation seems independent of the choice of the initial state. As we have seen in the previous two cases, the dominant character of the evolution is to drive the system towards a classical configuration for one of the modes. This classical configuration is identified as one whose Wigner function distribution is essentially a  function sharply  peaked around the classical trajectory. We have also seen that the coupling introduces a subleading correction to the sharply peaked nature of the distribution, giving the system a residual signature of its quantum origin. We will briefly describe the general nature of these results in the next section.

\section{Emergent classicality: a more general analysis}

The analysis of our  model with linear coupling turns out to be an important indicator of the role of the interaction in building up the correlation between the conjugate variables in a system which starts out quantum mechanically. Our choice of interaction might be useful in extending the analysis to other interesting cases such as free particle limit ($\omega_{1/2}\rightarrow0$), models of interaction of moduli field with normal matter field ($\omega_{1/2}\rightarrow i\omega_{1/2}$)\cite{Kiefer:2009zz, DeWitt} and so on. More specifically, we will be interested in the future extension of this work to gravitational degrees of freedom. We will be interested in examining the interaction of the scale factor of the universe (treated as a quantum system), which appears as the moduli field in quantum cosmological models, with the modes of gravitational radiation under this context. 

It is interesting to note  that (although in a drastically simplified form)  we have performed a full quantum analysis of the ``back-reaction" of one mode on another for this specific, simple coupling. However, the  description of the gravitational degrees of freedom in the quantum cosmological models are more complicated than the  linear coupling used here. Even then, if the role of interaction turns out to be  similar, we can hope for an understanding of the full back reaction and its role in emergent classicality of spacetime with a quantum 
matter field riding over it. It is also important to evaluate the relation between the coupling parameter and the residual quantum behavior in a more general context, which will be indicative of the scale at which the quantum attributes of spacetime, if any, will be important vis-\'{a}-vis the strength of the coupling parameter.

Another important aspect of classicalization in this scheme is the frequency dependence. We have seen that for an effective classicalization of mode $x_2$, its frequency should be vastly separated from that of the mode $x_1$, i.e. $f\gg 1$. In harmonic oscillator system, the frequency also determines the separation between the energy eigenstates. Thus the mode whose quantum levels are more separated, retains more of its quantum nature. It is not the energy of the system which is crucial; as we will see shortly, for the classicalization of mode $x_2$, it is not required to be in ground state. It can well be in an excited state or a combination of excited states. In the field theoretic generalization of this model we anticipate the density distribution of the corresponding field to determine the effective classicalization.
Now, as promised before, we will rapidly discuss the applicability of this model in a more general context.

\subsection{ Properties of the Propagator:}

In previous sections we have demonstrated the classicalization of a part of the system when the coupling turns super-critical if we start with a Gaussian initial state. However, similar results can be obtained for a more broad class including non-Gaussian states as well. For example we can 
prepare the system in an initial state
\bea
\Psi\{X_+,X_-;0\}\sim X_+\exp{\{-a X_+^2\}} \zeta(X_-)
\eea
where $\zeta(X_-)$ is an arbitrary function\footnote{We could, for example, work with the first excited state of the normal mode instead of
the ground state. This choice of initial state will include such configurations. First excited state of the coupled modes will just be a 
linear combination of such states.} of $X_-$ which is not odd and $a\in \mathbb{R}$. Computations for such states are analytically tractable and we obtain the Wigner
function of the system at late times to be:
\bea
W[X_+,P_+;X_-,P_-]
   \sim \left[4 a^2 \hbar ^2 (P_+-X_+ \Omega_+ \cot( \Omega_+ t))^2- \Omega_+^2 \left(a \hbar ^2 \cosec ^2( \Omega_+ t)-(P_+\cot( \Omega_+ t)+X_+ 
    \Omega_+)^2\right)\right]\nonumber\\
  \times   \exp{\left[-\frac{(P_+^2+\Omega_+^2 X_+^2)(\Omega_+^2+4 a^2 \hbar ^2)+(\text{$\Omega_+$}^2-4 a^2\hbar^2)\left((P_+^2-\Omega_+^2X_+^2)\cos(2\text{$\Omega_+$}t)+ 2P_+X_+\Omega_+\sin (2\text{$\Omega_+$}t)\right)}{4 a \text{$\Omega_+$}^2 \hbar ^2}\right]}\nonumber\\ \times   \delta(P_--\Omega_-X_-).
\eea
In the limit $a\rightarrow\Omega_+/2\hbar$, the $X_+$ part of the above Wigner function correctly reduces to that of the first excited state of the corresponding harmonic oscillator
$X_+$.
Once again, we can integrate one of the original modes to obtain the reduced Wigner function in the favor of the other one. We had already seen that the limit $f\gg\xi^2>1$ aligns $x_1,x_2$ with $X_+,X_-$. This corresponds to a limit $\theta\rightarrow \pi/2$
in \ref{Orthogonal2}. If we choose a large but finite $f$, it is equivalent to the choice of $\theta=\pi/2-\epsilon$ with $\epsilon\ll1$. In that case one can show that the reduced Wigner function for the mode $x_2$ for the case $a=\Omega_+/2\hbar$ is
\bea
{\cal W}_R[x_2,p_2]\approx \delta(p_2-\Omega_-x_2)+{\cal O}(\epsilon^2),
\eea
which again suggests that the mode $x_2$ is guided to a classical configuration during evolution.

Thus the results what we have obtained can be argued to be somewhat  more general. First, from the expression \ref{CCR11} we realize that  the Wigner function will be peaked on the classical trajectory upto arbitrary precision, for suitable choice of the parameters of the model.  Second, soon we will see that the  emergence of classicality is a consequence of the parameters in the model (i.e. the parameters in the Hamiltonian as well as those entering in the initial state) rather than that of the functional form of the initial state being Gaussian. To demonstrate this,  we can analyze the propagator of the theory. In the linear interaction model the propagator is the product of $X_+,X_-$ propagators,
\bea
{\cal G}_+[X_+,X_+']&\sim&\exp{\left[\frac{i \Omega_+}{2\hbar \sin{\Omega_+ t}}\{(X_+^2+X_+'^2)\cos{\Omega_+ t}-2 X_+ X_+'\}\right]} \nonumber\\
{\cal G}_-[X_-,X_-']&\sim&\exp{\left[\frac{i \Omega_-}{2\hbar \sinh{\Omega_- t}}\{(X_-^2+X_-'^2)\cosh{\Omega_- t}-2 X_- X_-'\}\right]}.
\eea
From above expressions we see that for an initial separable state which is not odd in $X_-$ (see Appendix C for details), the time evolved state will in general be
$$\Psi_t \sim \exp{\{ \frac{i\Omega_-}{2\hbar}X_-^2\}}\xi(X_+), $$
at late times, where $\xi(X_+)$ is a solely $X_+-$ dependent function determined by the initial state.
If we write the propagators in terms of original $x_1,x_2$ modes, we get
\bea
G(x_1,x_2)=\exp{\left[\frac{i \Omega_-}{2\hbar}\{(\cos^2{\theta} x_1^2+\sin^2{\theta} x_2^2 + \sin{2\theta}x_1x_2+ x_1'^2\cos^2{\theta}+\sin^2{\theta} x_2'^2 + \sin{2\theta}x_1'x_2')'\}\right]} \times \nonumber\\
\exp{\left[\frac{i \Omega_+}{2\hbar \sin{\Omega_+ t}}[(x_1^2+x_1'^2)\sin^2{\theta}+(x_2^2+x_2'^2)\cos^2{\theta}-\sin{2\theta}(x_1x_2+x_1'x_2')]\cos{\Omega_+ t}\right]} \nonumber\\
\exp{\left[-\frac{i \Omega_+}{\hbar \sin{\Omega_+ t}} (-\sin{\theta}x_1+\cos{\theta}x_2)(-\sin{\theta}x_1'+\cos{\theta}x_2')\right]}.
\eea
The limit $f\gg\xi^2$ implies $\theta \rightarrow \pi/2$. If $\theta=\pi/2 -\epsilon$ with a small $\epsilon$, we have 
\bea
G(x_1,x_2)={\cal G}_+[x_1,x_1']{\cal G}_-[x_2,x_2']\exp{\left[\frac{i \Omega_-}{\hbar}\{\epsilon(x_1x_2+x_1'x_2')\}\right]} \times \nonumber\\
\exp{\left[-
\frac{i \Omega_+}{\hbar}\{\epsilon(x_1x_2+x_1'x_2')\cot{\Omega_+t}-\epsilon(x_1x_2'+x_1'x_2)\cosec{\Omega_+t}\}\right]}, \label{GenProp}
\eea
up to the leading order in $\epsilon$. Now following the lines of discussion in Appendix C, we can see that for a general non-odd initial state, the time evolved state
will have certain interesting features. The integral of the propagator \ref{GenProp} over $x_2'$ will result
in a Gaussian state for $x_2$ at very late times (see the discussion in Appendix C). Thereafter any correction to $x_2$ part of the wave function can come only
from the last term in the second exponential when the propagator is integrated over $x_1'$. Therefore, we see that any departure from the state being a Gaussian in $x_2$, can come only from the last 
term in the second exponential. Rest of the  propagator gives a quadratic $x_2$ dependence in the exponential at late times. In fact, the amount of a
non-Gaussianity in $x_2$ is controlled by the parameter $\epsilon$ and the  $x_1$ dependence of the initial state. Clearly the dominant 
contribution in the (reduced) Wigner function (for $x_2$) will be a zero variance Gaussian at late times. A non zero variance comes from a second 
order ${\cal O}(\epsilon^2)$ correction to the $x_2$ dependent part in the time evolved state if the $x_1$ dependent part also happens to be a Gaussian
initially (for details refer to Appendix A). Any arbitrary initial dependence on  $x_1$  will contribute towards a subleading correction in the Wigner
function for $x_2$ giving it a residual quantum character as before.\\

\subsection{Robustness of the results}

Another interesting aspect of this model is its stability under the addition of more number of degrees of freedom with linear interactions. Again with a sufficiently strong coupling between just \textit{one pair} of modes we can turn one of the normal mode frequencies imaginary, so that if one of the original modes gets reasonably aligned with the corresponding normal mode, it acquires classicality. Further one can show,
appealing to the {\it Sylvester's criterion}\footnote{Sylvester's Criterion: A real symmetric matrix with positive diagonal elements is positive definite
if and only if each of its principal submatrix has positive determinant. Achieving criticality for just one pair denies the matrix the property of positive definiteness.}, that if such a configuration is achieved it will remain so, irrespective of the coupling parameter strength between other pairs. In that spirit, we can introduce an environment to the system without affecting the results, but in true sense that is not required for the classicalization in this particular context. Interestingly, if the
coupling parameters between different constituents are such that the majority of the eigenvalues are negative and the parameters of the theory are such 
that some of the initial modes get strongly projected along the corresponding normal modes, the evolution will guide the system to a configuration where
majority of its constituents embrace classical description while a subdominant part remains quantum (again this is not so difficult to see through the
propagator of the theory) sitting atop the classical background that has emerged spontaneously during evolution. This configuration is appealing because it has the 
potential to mimic the later stages of the evolution of the universe.

\section{Conclusions and Discussion}

In the present analysis, we worked with a  model of two  harmonic oscillators coupled linearly with each other. The aim of this work was to understand the role of back-reaction completely in an analytically solvable model, which could give us some insights in  more general contexts like cosmology and black hole evaporation. We studied the effect of the interaction on the evolution of the two quantum systems separately, in terms of reduced functions, when the initial state happens to be Gaussian in nature. The evolution  depends on the strength of coupling between the modes (systems) in an interesting way. If the coupling strength is below a critical value, the modes behave in a purely quantum manner devoid of any significant correlation in the canonically conjugate variables that appear in  the Wigner function. Further, we also obtain a thermal density matrix for one mode when the other mode is traced over. This thermal character arises without any  `environment' with large number of degrees of freedom acting as reservoir for the selected mode.

More interesting things start happening as soon as  the coupling parameter becomes larger than a critical value. Then, one of the modes (the one with the lower energy) spontaneously develops a significant classical correlation in a parameter range as revealed by the Wigner function, while the other mode still retains its quantum nature in the same sense.  For one of the modes, the reduced Wigner distribution function becomes sufficiently squeezed about the classical trajectory in the phase space and acquires a description which mimics a classical probability distribution very well. This appearance of spontaneous classicality is something which is independent 
of the choice of the initial state as the ground state of the coupled or the uncoupled modes. If we start with an initially uncoupled system in the ground state and then turn on a sufficiently strong coupling (non-perturbatively) then one of the modes (with a lower frequency)
gradually slips into a ``more classical" configuration while the other mode keeps behaving quantum mechanically, provided the two frequencies of the modes are sufficiently apart.

There is another surprising feature revealed by the analysis. Even when the coupling is super-critical and one mode acquires classicality, it does not become completely classical [i.e, the Wigner function does not becomes proportional to Dirac delta function at late times] but some degree of nonclassicality remains. This behavior is strengthened when we increase the coupling strength. So, if the coupling parameter in the super-critical regime gets stronger, the emergent classical features can get progressively diminished. Thus, the coupling parameter seems to be playing  a dual role. On the one hand, it makes one of the modes to get classical when it becomes  super-critical, while on the other hand it leads to  a residual quantum variance due to its coupling to the other mode which  remains quantum mechanical. For a given system, one can make out from the system parameters and the strength of the coupling whether or not there exists a regime of emergent classicality. 

This emergence of classicality  occurs for the smallest environment a system can get viz. in a system with two degrees of freedom of which one is traced out. Therefore, there is no role of decoherence in this behavior as can be verified from the reduced density matrix.
Furthermore,  the system which is pushed towards classical behavior also loses its thermal character in terms of reduced density matrix.

There have been attempts in the earlier literature  to obtain ``decoherence without decoherence'' in connection with the classicality of primordial fluctuations (like e.g in \cite{Kiefer:1998jk}). While our approach also achieves classicality without decoherence, there are significant differences between the two approaches. First, we \textit{crucially use the unboundedness} of the Hamiltonian which arises when the interaction is strong enough; as a result we have a tunable parameter which determines whether the relevant system will behave classically or not. Second, as we stressed earlier in this section, the system exhibits certain amount of nonclassicality even at late times which is residual effect arising from the interaction with the quantum system. Neither of these features are shared by any of the other approaches we are aware of suggesting that what we have here is a genuinely new mechanism. We also mention that our approach based on Wigner function is mathematically concrete and bypasses some of the (unsettled) conceptual issues related to superposition of states and whether diagonalisation of the density matrix (due to standard decoherence) is sufficient for classicality etc. We have tried to be as concrete as possible in our discussion. This helps us, for example, to study the behavior of the system at arbitrarily late times and verify that, classicality arising from this mechanism is \textit{not} a transient feature --- that is, the initial quantum nature will \textit{not} return after long Poincar\'{e} recurrence time.

The development of the classical behavior of one of the sub-systems is  practically independent of the choice of the initial state. Difference choices of  the initial state contribute  only towards the subleading departure of the system from developing a  classical character. In this sense, the classicalization of the subsystem is a robust result which is  initiated by a strong enough interaction. It will be  interesting to investigate if such a development is also present for more general interactions rather than just the linear coupling we have analyzed in this paper.

This role of interaction both as a messenger of classicality and a carrier of quantum uncertainty can have interesting implications for cosmology and black hole physics. A suitable generalization of this model will have the capacity of mapping the full quantum back reaction on the system of interest. One important avenue to try out the generalization of this model is in the context of cosmology where the back reaction from matter and gravitational perturbation degrees of freedom might help the background geometry to emerge classical on suitable scales decided by the interaction strength. Such a model may provide a picture of a fully quantum system  comprising of many different degrees of freedom (gravity and other fields) interacting with each other, gradually evolving to a system where quantum fields reside on an effectively classical background geometry. For instance, this will be a situation closely resembling the scenario for very early universe. We have also seen that it might not be necessary to constrain {\it all} the interactions of different matter fields with geometry. Only one of the coupling turning critical might well do the job, making the case more interesting from the point of view of plausibility.

If the results of the present analysis get just translated for the more general case, the strength of the interaction will also set a scale at which the true quantum properties of background geometry will be visible. We wish to study the applicability of this model on many different avenues (discussed throughout this paper) in a future work.

\section*{Acknowledgments}

The research of TP is partially supported by the J.C. Bose research grant of the Department of Science and Technology,
Government of India. The research of KP is supported by the Shyama Prasad Mukherjee fellowship of the Council of Scientific and Industrial Research, Government of India.

\begin{appendices}
\section{: General Wigner Function for supercritical case}
We consider a general Gaussian initial state, so as to be able to reach up to various cases we discussed previously
\bea
\Psi(X_-,X_+;0)=N \exp{\{-(\alpha X_-^2+\tilde{\alpha} X_+^2 + 2 \beta X_+ X_-)\}}.\label{Initial1App}
\eea
Evolution of this general Gaussian state will be obtained from the Kernel of the system, obtained easily when written in normal modes
\bea
\psi(X_-,X_+;t)=\int dX_-'dX_+'G(X_-,X_-',X_+,X_+';t,0)\psi(X_-',X_+';0), \label{TE1App}
\eea
where, 
\bea
G(X_-,X_-',X_+,X_+';t,0)= \left(\frac{\Omega_+}{2\pi i \hbar \sin{\Omega_+ t}}\right)^{\frac{1}{2}}
\left(\frac{\Omega_-}{2\pi i \hbar \sinh{\Omega_- t}}\right)^{\frac{1}{2}}\times{\cal G}_+\times{\cal G}_-,
\eea
with, 
\bea
{\cal G}_+&=&\exp{\left[\frac{i \Omega_+}{2\hbar \sin{\Omega_+ t}}\{(X_+^2+X_+'^2)\cos{\Omega_+ t}-2 X_+ X_+'\}\right]}, \nonumber\\
{\cal G}_-&=&\exp{\left[\frac{i \Omega_-}{2\hbar \sinh{\Omega_- t}}\{(X_-^2+X_-'^2)\cosh{\Omega_- t}-2 X_- X_-'\}\right]}. \label{Kernel}
\eea
For the initial state of the above mentioned kind, the time evolution gives
\bea
\Psi(X_-,X_+;t)=N\frac{\sqrt{\alpha_1\alpha_2}}{\sqrt{\tilde{\alpha}k_1-\beta^2-k_1\alpha_1\beta_1}}\exp{\{\alpha_1\beta_1X_+^2+\alpha_2\beta_2X_-^2+
\frac{\alpha_2^2}{\alpha-\alpha_2\beta_2}X_-^2\}}\nonumber\\
\times \exp{\{ \frac{(k_1\alpha_1X_++\beta\alpha_2X_-)^2}{k_1(\tilde{\alpha}k_1-\beta^2-k_1\alpha_1\beta_1)}\}}, \label{TimeEvolvedApp}
\eea
with the condition of the state being normalizable imposed through
$$ Re[\frac{\beta^2}{k_1}+\alpha_1\beta_1-\tilde{\alpha}]\leq0,$$
where we have 
\bea
\alpha_1&=&\frac{i\Omega_+}{2\hbar\sin{\Omega_+t}}; \hspace{0.55 in} \beta_1=\cos{\Omega_+t} \nonumber\\
\alpha_2&=&\frac{i\Omega_-}{2\hbar\sinh{\Omega_-t}}; \hspace{0.475 in} \beta_2=\cosh{\Omega_-t}\nonumber\\
k_1&=&\alpha-\alpha_2\beta_2.
\eea
Now, the time evolved state \ref{TimeEvolvedApp} can be rewritten in a more comprehensible form as
\bea
\Psi(X_-,X_+;t)=N\frac{\sqrt{\alpha_1\alpha_2}}{\sqrt{\tilde{\alpha}k_1-\beta^2-k_1\alpha_1\beta_1}}\exp{\{-(B X_+^2+C X_-^2 + 2 D X_+X_-)\}}, \label{PsiTApp}
\eea
where, we have
\bea
B &=& -\left[\frac{i\Omega_+}{2\hbar}\cot{\Omega_+t}-\frac{(\alpha-\frac{i\Omega_-}{2\hbar}\coth{\Omega_-t})
\frac{\Omega_+^2}{(2\hbar\sin{\Omega_+t})^2}}{\alpha\tilde{\alpha}-\beta^2-\frac{\Omega_+\Omega_-}{4 \hbar^2}\coth{\Omega_-t}\cot{\Omega_+t}-i(\frac{\Omega_-\tilde{\alpha}}{2\hbar}\coth{\Omega_-t}+
\frac{\Omega_+\alpha}{2\hbar}\cot{\Omega_+t})} \right],\\
D &=& -\left[\frac{\beta \frac{\Omega_+\Omega_-}{4 \hbar^2}\cosech{\Omega_-t}\cosec{\Omega_+t} }{\alpha\tilde{\alpha}-\beta^2-\frac{\Omega_+\Omega_-}{4 \hbar^2}\coth{\Omega_-t}\cot{\Omega_+t}-i(\frac{\Omega_-\tilde{\alpha}}{2\hbar}\coth{\Omega_-t}+
\frac{\Omega_+\alpha}{2\hbar}\cot{\Omega_+t})}\right],\\
C &=& -\left[\frac{i\frac{\Omega_-\alpha}{2\hbar}\coth{\Omega_-t}+ \frac{\Omega_-^2}{4 \hbar^2}}{\alpha-i\frac{\Omega_-}{2\hbar}\coth{\Omega_-t}}+
\frac{\beta^2\alpha_2^2}{k_1(\tilde{\alpha}k_1-\beta^2-k_1\alpha_1\beta_1)}\right].
\eea
This is the most generic form of the evolution of \ref{Initial1App}. From here we can easily obtain the late time limits of this state and obtain the corresponding Wigner function, which is given in \ref{GeneralWF}. We see that coefficient $D$ asymptotically gets suppressed and we essentially have a decoupled state in normal modes. The reduced Wigner function for modes $x_1$ and $x_2$ are given in the Morikawa form with parameters
\bea
\bar{\gamma}_1^2 &=&\frac{(B_R\cos^2{\theta}+C_R\sin^2{\theta})}{2\{(B_R-C_R)^2+(B_I-C_I)^2\}\sin^2{\theta}\cos^2{\theta}-2B_RC_R}\nonumber\\
\bar{\alpha}^2&=&\frac{2B_RC_R\left[\{(B_R-C_R)^2+(B_I-C_I)^2\}\sin^2{2\theta}+4B_RC_R\right]}{[2\{(B_R-C_R)^2+(B_I-C_I)^2\}\sin^2{\theta}\cos^2{\theta}-2B_RC_R](B_R\cos^2{\theta}+C_R\sin^2{\theta})}\nonumber \\
\bar{\beta} &=& \frac{2(B_RC_I\cos^2{\theta}+C_R B_I\sin^2{\theta})}{(B_R\cos^2{\theta}+C_R\sin^2{\theta})}
\eea
and for mode $x_2,$
\bea
\bar{\gamma}_2^2 &=&\frac{(B_R\sin^2{\theta}+C_R\cos^2{\theta})}{2\{(B_R-C_R)^2+(B_I-C_I)^2\}\sin^2{\theta}\cos^2{\theta}-2B_RC_R}. 
\eea

Clearly for the super-critical limit $C_R \rightarrow 0$ at late times, makes the parameter $\bar{\alpha}$ vanish asymptotically for both the variables. Furthermore we can verify that coefficients $B_R,B_I$ and $C_I$ have strength of typical quantum systems. The interesting range $f\gg\xi^2$ either keeps $\bar{\gamma}$ around the value of the corresponding quantum value (for $x_1$) or makes it much smaller than it (for $x_2$).

\section{: General Wigner Function for sub-critical case}

The time evolution for the sub-critical case is governed by the standard normal mode propagators satisfying harmonic oscillator equations of motion classically. With a state of type \ref{Initial1App} and the propagators
\bea
{\cal G}_+&=&\exp{\left[\frac{i \Omega_+}{2\hbar \sin{\Omega_+ t}}\{(X_+^2+X_+'^2)\cos{\Omega_+ t}-2 X_+ X_+'\}\right]} \nonumber\\
{\cal G}_-&=&\exp{\left[\frac{i \Omega_-}{2\hbar \sin{\Omega_- t}}\{(X_-^2+X_-'^2)\cos{\Omega_- t}-2 X_- X_-'\}\right]},
\eea
we obtain the a state of type \ref{PsiTApp} whose coefficients keep oscillating in this case without having any meaningful notion of an asymptotic value
\bea
B &=&-\left[\frac{i \Omega_+}{2\hbar}\cot{\Omega_+t}-\frac{(\alpha-\frac{i\Omega_-}{2\hbar}\cot{\Omega_-t})\frac{\Omega_+^2}{h\hbar^2}\cosec{\Omega_+t}}
{\alpha\tilde{\alpha}-\beta^2-\frac{\Omega_+\Omega_-}{4 \hbar^2}\coth{\Omega_-t}\cot{\Omega_+t}-i(\frac{\Omega_-\tilde{\alpha}}{2\hbar}\coth{\Omega_-t}+
\frac{\Omega_+\alpha}{2\hbar}\cot{\Omega_+t})} \right],\nonumber\\
D &=& -\left[\frac{\beta \frac{\Omega_+\Omega_-}{4 \hbar^2}\cosec{\Omega_-t}\cosec{\Omega_+t} }{\alpha\tilde{\alpha}-\beta^2-\frac{\Omega_+\Omega_-}{4 \hbar^2}\coth{\Omega_-t}\cot{\Omega_+t}-i(\frac{\Omega_-\tilde{\alpha}}{2\hbar}\cot{\Omega_-t}+
\frac{\Omega_+\alpha}{2\hbar}\cot{\Omega_+t})}\right],\nonumber\\
C &=& -\left[\frac{i\frac{\Omega_-\alpha}{2\hbar}\cot{\Omega_-t}- \frac{\Omega_-^2}{4 \hbar^2}}{\alpha-i\frac{\Omega_-}{2\hbar}\cot{\Omega_-t}}+
\frac{\beta^2\alpha_2^2}{k_1(\tilde{\alpha}k_1-\beta^2-k_1\alpha_1\beta_1)}\right]. 
\eea
These general expressions for the parameters can be extended analytically to the super-critical case by transformation
$\Omega_-\rightarrow i\Omega_-$. Thus the most general results can be obtained from this transformation for super-critical case as well.
We can rewrite the Wigner function \ref{GeneralWF} in terms of $x_1$ and $x_2$ modes as
\bea
W(x_1,x_2,p_1,p_2)=|A|^2\frac{2\pi}{\sqrt{B_RC_R-D_R^2}}\exp{\{-(\alpha_1 x_1^2+\alpha_2x_2^2+\beta x_1 x_2+\sigma_1p_1^2+\sigma_2p_2^2+
\lambda p_1p_2)\}}\nonumber\\
\times\exp{\{-(\gamma_1x_1p_1+\gamma_2x_2p_2+\delta_1 x_1p_2+\delta_2x_2p_1)\}},
\eea
where,
\bea
\alpha_1&=&\frac{2}{B_RC_R-D_R^2}[\{B_R(C_R^2+C_I^2)-2D_ID_RC_I+C_R(D_I^2-D_R^2)\}\cos^2{\theta}\nonumber\\
&+&\{B_I^2C_R+B_R^2C_R-2D_ID_RB_I+B_R(D_I^2-D_R^2)\}\sin^2{\theta}\nonumber\\
&+&\{-(B_RC_I+B_IC_R)D_I+(B_IC_I-B_RC_R+D_I^2)D_R+D_R^3\}\sin{2\theta}], \\
\nonumber\\
\alpha_2&=&\frac{2}{B_RC_R-D_R^2}[\{B_I^2C_R+B_R^2C_R-2D_ID_RB_I+B_R(D_I^2-D_R^2)\}\cos^2{\theta},\nonumber\\
&+&\{B_R(C_R^2+C_I^2)-2D_ID_RC_I+C_R(D_I^2-D_R^2)\}\sin^2{\theta},\nonumber\\
&+&\{(B_RC_I+B_IC_R)D_I-(B_IC_I-B_RC_R+D_I^2)D_R-D_R^3\}\sin{2\theta}], \\
\beta &=&\frac{2}{B_RC_R-D_R^2}[2\{(B_RC_I+B_IC_R)D_I-(B_IC_I-B_RC_R+D_I^2)D_R-D_R^3\}\cos{2\theta},\nonumber\\
        &+&\{-B_I^2C_R-B_R^2C_R+2(B_I-C_I)D_ID_R+C_R(D_I^2-D_R^2)+B_R(C_R^2+C_I^2-D_I^2+D_R^2)\}\sin{2\theta}],\nonumber\\
     \\
\lambda&=&\frac{-D_R\cos{2\theta}+(B_R-C_R)\sin{\theta}\cos{\theta}}{B_RC_R-D_R^2}, \\
      \nonumber \\
\gamma_1&=&  \frac{B_RC_I+B_IC_R-2D_ID_R+(B_RC_I-B_IC_R)\cos{2\theta}+\{-(B_R+C_R)D_I+(B_I+C_I)D_R \}\sin{2\theta}}{B_RC_R-D_R^2},  \\     
\gamma_2&=&  \frac{B_RC_I+B_IC_R-2D_ID_R-(B_RC_I-B_IC_R)\cos{2\theta}-\{-(B_R+C_R)D_I+(B_I+C_I)D_R \}\sin{2\theta}}{B_RC_R-D_R^2}, \nonumber\\
       \\ 
\delta_1&=&\frac{2[(C_RD_I-C_ID_R)\cos^2{\theta}+(B_RC_I-B_IC_R)\cos{\theta}\sin{\theta}+(B_ID_R-B_RD_I)\sin^2{\theta}]}{B_RC_R-D_R^2},\\
\nonumber\\
\delta_2&=&\frac{2[(B_RD_I-B_ID_R)\cos^2{\theta}-(C_RB_I-C_IB_R)\cos{\theta}\sin{\theta}+(C_ID_R-C_RD_I)\sin^2{\theta}]}{B_RC_R-D_R^2}, \\
             \nonumber \\
\sigma_1&=& \frac{B_R\cos^2{\theta}+C_R\sin^2{\theta}+D_R\sin{2 \theta}}{2(B_RC_R-D_R^2)}, \\
\nonumber\\
\sigma_2&=& \frac{C_R\cos^2{\theta}+B_R\sin^2{\theta}-D_R\sin{2 \theta}}{2(B_RC_R-D_R^2)}.
\eea
This expression is true in general for both the supercritical as well as the sub-critical case, only for the fact that in the supercritical case some of the parameters become much simplified.
We obtain the expression for the reduced Wigner function for the general case by tracing over the $x_1$ mode.
\bea
{\cal W}_R(x_2,p_2)=\frac{2\pi}{\sqrt{4\alpha_1\sigma_1-\gamma_1^2}}\exp{\left[-\left(\sigma_2-\frac{\delta_1^2}{4\alpha_1}+\frac{4\alpha_1(\gamma_1\delta_1-2\alpha_1\lambda)^2}{\gamma_1^2-4\alpha_1\sigma_1} \right)p_2^2 \right]}\times \nonumber\\
\exp{\left[-\left(\delta_2-\frac{\sigma_1\beta}{2\alpha_1}+\frac{8\alpha_1(\gamma_1\delta_1-2\alpha_1\lambda)(\beta\gamma_1-2\alpha_1\gamma_2)}{\gamma_1^2-4\alpha_1\sigma_1} \right)p_2x_2 -\left(\alpha_2-\frac{\beta^2}{4\alpha_1}+ \frac{4\alpha_1(\beta\gamma_1-2\alpha_1\gamma_2)^2}{\gamma_1^2-4\alpha_1\sigma_1} \right)x_2^2\right]}. \nonumber\\
\eea
The variance in the correlation trajectory will now be determined by the inverse of the coefficient of $p_2^2$ in the above expression. Thus, we have
$$\bar{\gamma}^2=\frac{1}{\left(\sigma_2-\frac{\delta_1^2}{4\alpha_1}+\frac{4\alpha_1(\gamma_1\delta_1-2\alpha_1\lambda)^2}{\gamma_1^2-4\alpha_1\sigma_1} \right)}.$$
We can verify that for the supercritical limit the region $f\gg\xi^2$ sends the variance to the vanishingly small values asymptotically, whereas the sub-critical case does not contain any such regime. We can also obtain the results for the initially uncoupled state results, which is the case ($D=0$), from here.

\section{: Wigner function for an arbitrary initial state for $X_-$ }

For the inverted mode $X_-$ we have the expression for the Kernel in \ref{Kernel}. We will try to obtain the Wigner function for a general initial state
$\Psi_0(X_-)$ which is not vanishing at $X_-=0$. The time evolved state for this is given as
\bea
\Psi_t(X_-)=\int dX_-'{\cal G}_-(X_-,X_-')\Psi_0(X_-').
\eea
Since, $\Psi_0(X_-)$ is a continuous differentiable function, we can Taylor expand it around $X_-=0$ as
\bea
\Psi_0(X_-)=\Psi_0(0)+\sum_{n=1} c_n X_-^n, \label{ExPofGenX_}
\eea
where $c_n=\Psi^{(n)}(X_-)|_0$. Therefore, the time evolution will be of the type
\bea
\Psi_t(X_-)= \exp{\left(\frac{i\Omega_-}{2\hbar}X_-^2\right)}\left[c_0+\sum_n c_n \int dX_-'X_-^{'n}\exp{[i\tilde{a}X_-^{'^2}-i \tilde{b}X_-X_-']}\right],
 \label{GenX_}
\eea
where $\tilde{a}=\frac{\Omega_-}{2\hbar}\coth{\Omega_- t}$ and
      $\tilde{b}=\frac{\Omega_-}{\hbar \sinh{\Omega_- t}}$.
We obtain the time evolved state, using 
\bea
\int dx x^{n}\exp{[i\tilde{a}x^2-i \tilde{b}x y]}=-\frac{1}{2}\tilde{a}(-i \tilde{a})^{-\frac{n}{2}-\frac{5}{2}} \times \left[\sqrt{-i \tilde{a}} \tilde{b}
   \left(1-(-1)^n\right) y \Gamma \left(\frac{n}{2}+1\right) \,
   _1F_1\left(\frac{n}{2}+1;\frac{3}{2};-\frac{i \tilde{b}^2 y^2}{4 \tilde{a}}\right)\right]\nonumber\\
-\frac{1}{2}\tilde{a}(-i \tilde{a})^{-\frac{n}{2}-\frac{5}{2}} \times\left[\tilde{a}  
   \left((-1)^n+1\right) \Gamma \left(\frac{n+1}{2}\right) \,
   _1F_1\left(\frac{n+1}{2};\frac{1}{2};-\frac{i \tilde{b}^2 y^2}{4 \tilde{a}}\right)\right],
\eea
where $\,   _1F_1(a,b,z) $ is the Kummer confluent hypergeometric function and $\Gamma(z) $ is the Gamma function. We realize that the parameter
$\tilde{b}$ becomes vanishingly small at late times and one can verify that in that limit the series expansion of \ref{GenX_} will have 
vanishing odd terms.
The general structure of the time evolved state, however, will be
\bea
\Psi_t(X_-)= \exp{\left(\frac{i\Omega_-}{2\hbar}X_-^2\right)}\left[\alpha_0+\sum_n \alpha_n\tilde{b}^{n} X_-^{2n} \right],
\eea
where $\alpha_0$ receives contribution from all the even terms of the expansion of \ref{ExPofGenX_} and is non-zero generically.
Therefore, the Wigner function will assume the structure 
\bea 
W[X_-,P_-]\sim |\alpha_0|^2\delta(P_--\Omega_-X_-) + \sum_n \tilde{b}^{n}f_n(X_-,P_-).
\eea
Therefore, we see that unless the initial state is odd, the Wigner function at very late times is dominated by the Dirac delta distribution (upto a normalization).

\section{: Classical equations of trajectories for coupled modes}

The mode $x_2$ is a linear combination of uncoupled modes $X_+$ and $X_-$. Now, $X_{+}$ is a usual harmonic oscillator and hence $X_{+}$ and $P_{+}$ are bounded. But $X_{-}$ and $P_{-}$ will grow without bound.  Therefore, $x_{2}$ and $p_{2}$, has the time dependent behavior as,
\begin{align}
x_2 &= \sin \theta (A_{-} e^{ \Omega_- t} + B_{-} e^{- \Omega_- t}) + \cos \theta (A_{+} e^{i \Omega_+ t} + A_{+} e^{-i \Omega_+ t} ) \\ 
p_2 &= \Omega_- \sin \theta \left(A_{-} e^{ \Omega_- t} - B_{-} e^{- \Omega_- t}\right) + i  \Omega_+ \cos \theta\left( A_{+} e^{i \Omega_+ t} - A_{+} e^{-i \Omega_+ t}\right).
\end{align}
At late times, we may neglect the contribution of oscillatory terms (specially in the scenario when $x_2$ is aligned dominantly along $X_-$) to approximate the trajectory as
\begin{equation} \label{x2_trajec}
 p_{2}-\Omega_- x_2 \approx 0~. 
\end{equation}
Note that this becomes precise in the limit where $\cos \theta$ tends to zero, which is a limit that we shall consider later in the analysis. A similar analysis can be done for the $x_1$ mode gives qualitatively similar results.
\end{appendices}

\end{document}